\documentclass[a4paper,12pt]{article}
\pdfoutput=1
\usepackage[margin=1in]{geometry}
\usepackage{graphicx,xcolor}
\usepackage{mathtools,braket,blkarray}
\allowdisplaybreaks
\usepackage{amssymb,amsfonts,bbm,relsize}
\usepackage[width=0.9\textwidth,footnotesize]{caption}
\usepackage[font=scriptsize]{subcaption}
\usepackage{cite,hyperref,url}
\hypersetup{pageanchor=false}
\usepackage{booktabs,multirow,makecell}
\usepackage[utf8]{inputenc}
\usepackage{cleveref}
\usepackage{amsmath}

\newcommand{\be}{\begin{equation}}
\newcommand{\ee}{\end{equation}}
\newcommand{\ba}{\begin{eqnarray}}
\newcommand{\ea}{\end{eqnarray}}

\begin{document}

\begin{titlepage}
\begin{center}
{\bf\Large Flipped SU(5): unification, proton decay, fermion masses and gravitational waves
  } \\[12mm]
Stephen~F.~King$^{a}$~\footnote{E-mail: \texttt{king@soton.ac.uk}},
George~K.~Leontaris$^{b}$~\footnote{E-mail: \texttt{leonta@uoi.gr}},
Ye-Ling~Zhou$^{c}$~\footnote{E-mail: \texttt{zhouyeling@ucas.ac.cn}}
\\[-2mm]

\end{center}
\vspace*{0.50cm}
\centerline{$^{a}$ \it
School of Physics and Astronomy, University of Southampton,}
\centerline{\it
SO17 1BJ Southampton, United Kingdom }
\centerline{$^{b}$~\it
			Physics Department, University of Ioannina, 45110, Ioannina, 	Greece}
 \vspace*{0.2cm}
\centerline{$^{c}$ \it School of Fundamental Physics and Mathematical Sciences,}
\centerline{\it Hangzhou Institute for Advanced Study, UCAS, Hangzhou 310024, China}
\vspace*{1.20cm}

\begin{abstract}
{\noindent
We study supersymmetric (SUSY) flipped $SU(5)\times U(1)$  unification, focussing on its predictions for proton decay, fermion masses and gravitational waves. We performed a two-loop renormalisation group analysis and showed that the SUSY flipped $SU(5)$ model predicts a high GUT scale $M_{\rm GUT} > 10^{16}$ GeV. 
We also investigated  the restrictions on the $M_{B-L}$ scale which is associated with the $U(1)_{\chi}$ breaking scale.
We found that the $M_{B-L}$ scale can vary in a broad region with negligible or little effect on the value of $M_{\rm GUT}$. 
Proton decay in this model is induced by dimension-6 operators only.
The dimension-5 operator induced by SUSY contribution is suppressed due to the missing partner mechanism. 
We found that the partial decay width $p \to \pi^0 e^+$ is high suppressed, being at least one order of magnitude lower than the future Hyper-K sensitivity. We also studied fermion (including neutrino) masses and mixings which can also influence proton decay.
We presented two scenarios of flavour textures to check the consistency of the results with fermion masses and mixing. 
The $B-L$ gauge breaking leads to the generation of cosmic strings. The $B-L$ scale here is not constrained by gauge coupling unification. 
If this scale is very close that of GUT breaking, strings can be unstable due to the decay to monopole-antimonople pair. Such metastable strings can be used to explain the NANOGrav signals
of stochastic gravitational wave background, which may be interpreted here as resulting from the decay of metastable cosmic strings.

}
\end{abstract}
\end{titlepage}

\section{Introduction}

For more than half a century, since Georgi and Glashow proposed $SU(5)$ \cite{Georgi:1974sy},
numerous Grand Unified Theories (GUTs) have been constructed with the goal of reproducing the stupendous success of the standard model and, at the same time, providing new testable (falsifiable) predictions for particle physics and cosmology. In the years that followed, several new experimental facilities were conceived and built to verifying or disproving  predictions,  and imposing bounds on various exotic processes  such as proton decay, and flavour changing reactions. 
Furthermore, accumulating cosmological data, and in particular the recent announcements of  pulsar timing array (PTA) experiments such as the 
NANOGrav~\cite{NANOGrav:2023gor}, EPTA~\cite{EPTA:2023fyk}, PPTA \cite{Reardon:2023gzh} and CPTA \cite{Xu:2023wog} collaborations,
providing  strong evidence of low frequency stochastic gravitational background can find interpretations through decays of cosmic strings \cite{NANOGrav:2023hvm} whose creation and subsequent evolution are intertwined with the symmetry breaking scales of novel  $U(1)$'s and GUTs in general.

Among the most promising  unified theories which have been proposed to embed the Standard Model is the $SU(5)\times U(1)_{\chi}$, the so called  flipped $SU(5)$~\cite{Barr:1981qv, Derendinger:1983aj}, where the hypercharge generator  is a linear combination of the $U(1)\in SU(5)$ and $ U(1)_{\chi}$. 
In contrast to the Georgi-Glashow $SU(5)$,  spontaneous symmetry breaking in flipped $SU(5)$ requires only a Higgs pair of fundamentals  ${\bf 10}+\overline{
\bf 10}$, and thus, it  
dispenses with the use of the Higgs adjoint (i.e., ${\bf 24}\in SU(5)$)  representation. 
Although this is not a unified gauge model per se,  yet it is a  subgroup of the  unified $SO(10)$ GUT and, as such, it possesses  a number of interesting features that make it quite compelling. 
In string theory, for example, mechanisms such as  Wilson lines or fluxes can be  implemented for the $SO(10)$ GUT  symmetry  breaking, so that the effective theory particle spectrum is essentially described by the flipped $SU(5)$ model.  Moreover,  its supersymmetric (SUSY) version has been one of the few  admissible   candidates in particular heterotic string and F-theory constructions which do not  provide adjoint or larger  Higgs  representations to trigger the spontaneous symmetry breaking.
Flipped $SU(5)$ is also attractive since the unwanted colour triplets which accompany the Higgs doublets of GUT models, can more easily be given large masses via the missing partner mechanism. This has the additional benefit of suppressing all dimension-5 operators contributing to baryon violating processes. Thus, proton decay  proceeds mainly through dimension-6 operators, with only a smaller number of operators contributing thereby suppressing the proton decay rate.
Recently we have studied flipped $SU(5)$
in the framework of modular symmetry~\cite{Charalampous:2021gmf},
focussing on the origin of fermion masses, mixing and hierarchies, including the neutrino spectrum which can arise from a type II seesaw mechanism.
For some recent works in the literature, see e.g., \cite{Ellis:2020qad, Mehmood:2020irm, Antoniadis:2020txn, Abid:2021jvn, Ellis:2021vpp, Antoniadis:2021rfm, Lazarides:2023iim, Basilakos:2023jvp}.

In this paper we shall focus on the 
experimental tests of SUSY flipped $SU(5)$, including unification, proton decay, fermion masses and gravitational waves which arise from cosmic string loops associated with the breaking of the high energy $U(1)_{B-L}$.
The study here complements previous recent studies where such testable predictions have been studied in the framework of $SO(10)$
\cite{Fu:2023mdu,Fu:2022lrn,King:2021gmj,King:2020hyd}. Unlike the case of SUSY $SO(10)$ \cite{Fu:2023mdu}, we shall find that the proton decay rate is highly suppressed, rendering it practically unobservable in the forseeable future. Moreover, the SUSY breaking scale and the $U(1)_{B-L}$ breaking scales are essentially free parameters in flipped $SU(5)$, allowing for a wide range of gravitational wave signatures. This freedom allows the NANOGrav data to be fitted more easily in SUSY flipped $SU(5)$ than in SUSY $SO(10)$, for high $B-L$ breaking scales where unstable strings are possible, due to their decay into monopole-antimonopole pairs, without violating high frequency LIGO bounds. By contrast, in SUSY $SO(10)$, proton decay places a severe constraint on this scenario, requiring a split SUSY spectrum \cite{Fu:2023mdu}, which is not necessary for SUSY flipped $SU(5)$ where any SUSY spectrum is possible.

The layout of the remainder of this paper is as follows.	In section 2 we present the spectrum of the model and describe the  breaking  pattern of the (flipped) $SU(5)\times U(1)_{\chi}$ symmetry 	which involves  the sequence of scales $M_{\rm GUT} > M_{B-L} > M_{SUSY}$.
In section 3 we perform a two-loop renormalisation group analysis for the gauge couplings  running from the GUT scale (assumed to be the  scale where the $g_3$  and $g_2$ couplings attain a common value) down to the electroweak scale, and investigate  the restrictions on the $M_{B-L}$ scale which is associated with the $U(1)_{\chi}$ breaking.
It is found that the $M_{B-L}$ scale can vary in a broad region between $M_{\rm GUT}$  and $M_{SUSY}$  with negligible or little  effect on the value of $M_{\rm GUT}$ which is found to be $M_{\rm GUT}\gtrsim 10^{16}$ GeV. Next, in section 4 we discuss  proton decay.
We specifically show  that the predictions for the prevalent decay rate for $p\to \pi^0e^+$ of this model is practically unobservable by future super-K experimental measurements. Moreover,  in section 5 we discuss proton decay in conjunction with specific textures  of the fermion mass matrices with particular emphasis on the role  of the lepton mixing matrix.
In section 6 we investigate the  possibility of interpreting the 
recently reported data on stochastic gravitational wave background emissions through the decay of metastable cosmic strings produced due to  spontaneous breaking of $U (1)_{\chi}$  at a high-scale. In section 7  we present our conclusions.

\section{The framework}

The Flipped  $SU(5)$ model ~\cite{Barr:1981qv, Derendinger:1983aj} is based on the $SU(5)\times U(1)_{\chi}$  gauge symmetry. It was reconsidered as a possible superstring alternative to Georgi-Glashow $SU(5)$ due to  the fact that its spontaneous breaking to SM symmetry requires only a pair of ${\bf 10}+\overline{\bf 10}$ Higgs representations and does not need any adjoint Higgs representation.

The spontaneous breaking of the flipped $SU(5)$ symmetry  to the SM occurs via the following chain 
\begin{eqnarray}
&G^{\rm flip, S}_{51} \equiv SU(5) \times U(1)_\chi \times {\rm SUSY} \nonumber\\
&\hspace{5mm}{({\bf 24},0)}~\Big\downarrow~\text{broken at} ~ M_{\rm GUT} \nonumber\\
&G^{\rm S}_{3211} \equiv SU(3)_c \times SU(2)_L \times U(1)_y \times U(1)_\chi  \times {\rm SUSY} \nonumber\\
&\hspace{30mm}{(\overline{\bf 10},- \frac12) \supset {({\bf 1},{\bf 1}, 1, - \frac12)}}~\Big\downarrow~\text{broken at} ~ M_{B-L} \nonumber\\
&G_{\rm MSSM} \equiv SU(3)_c \times SU(2)_L \times U(1)_Y \times {\rm SUSY} \,, \nonumber\\
& \downarrow~ \nonumber\\
&G_{\rm SM} \equiv SU(3)_c \times SU(2)_L \times U(1)_Y \,.
\end{eqnarray}
For abbreviation, we denote the lowest and the second lowest symmetry above the symmetry of the minimal supersymmetric model $G_{\rm MSSM} \equiv SU(3)_c \times SU(2)_L \times U(1)_Y \times {\rm SUSY}$ as $G^{\rm S}_{3211}$ and $G_{51}^{\rm flip,S}$, where the superscript $^{\rm S}$ means the SUSY is conserved. In the symmetry $G^{\rm S}_{3211}$, the mixing of two $U(1)$ symmetries is just a re-organisation of $U(1)_Y$ and $U(1)_{B-L}$, $U(1)_y \times U(1)_\chi \simeq U(1)_Y \times U(1)_{B-L}$.
The hypercharge $Y$ and the gauged $B-L$ number are both linear combinations of the $y$ and $\chi$,
\begin{eqnarray}
Y&=&-\frac 15\left( y+ 2\chi \right) \,, \nonumber\\
B-L &=& \frac25\left( 2y- \chi \right)~.
\end{eqnarray}
We denote scales for $G^{\rm S}_{3211}$ breaking (essentially the $U(1)_{B-L}$ breaking) and $G_{51}^{\rm flip,S}$ breaking (essentially the $SU(5)$ breaking) as $M_{B-L}$ and $M_{\rm GUT}$, respectively. Representations of Higgs multiplets achieving the breaking are given on the left hand side of the arrow.
The spontaneous breaking of $G_{51}^{\rm flip,S} \to G^{\rm S}_{3211}$ is achieved via a ${\bf 24}$-plet of $SU(5)$ which is neutral in $U(1)_\chi$. This multiplet includes a trivial singlet of $G^{\rm S}_{3211}$. It gains a non-zero VEV, leading to the breaking of $SU(5) \to SU(3)_c \times SU(2)_L \times U(1)_y$.
The breaking of $U(1)_{B-L}$ is achieved by a $(\overline{\bf 10}, -\frac12)$ of $SU(5) \times U(1)_\chi$ which includes a colour- and electroweak-singlet scalar with charges $1$ and $-\frac12$ in $U(1)_y$ and $U(1)_\chi$, respectively, charge 1 in $U(1)_{B-L}$, and neutral in $U(1)_Y$.

The particle content is shown in Table~\ref{tab:particle_content}. Here, we included particles by following the economical hypothesis that only a minimal content is included to be consistent with the SM and neutrino masses, and to achieve the symmetry breaking. In the fermion
sector, only the SM fermions and right-handed neutrinos will be considered, all arranged as ${\bf 10}$, $\overline{\bf 5}$ and ${\bf 1}$ of $SU(5)$.
In the Higgs sector, only those Higgses used to generate fermion masses and achieve the
GUT symmetry breaking will be presented. 

\begin{table}[h!]
\begin{center}
\begin{tabular}{|c | c | l |}
\hline \hline
 & Multiplet & Role in the model
\\\hline
 &$({\bf 10},-\frac{1}{2})$ & Decomposed to $Q, d^c, \nu^c$ \\
Fermions &$(\bar{\bf 5}, +\frac{3}{2})$ & Decomposed to $u^c$ and $L$ \\
 &$({\bf 1}, -\frac{5}{2})$ & Identical to $e^c$ \\\hline
 &$({\bf 5}, +1)$ & Generating Dirac masses for fermions \\
 &$(\overline{\bf 10},{+\frac 12})$ & Generating $\nu^c$ mass and triggers $U(1)_{B-L}$ breaking \\
 Higgs &$(\overline{\bf 5}, -1)$ & For anomaly free \\
 &$({\bf 10}, -\frac12)$ & For anomaly free
 \\

 & $({\bf 24},0)$ & triggers the breaking of $SU(5)$ \\
 \hline \hline
\end{tabular}
\end{center}
\caption{{The $SO(10)$ representations of the fields (including matter and Higgses) %and Wilson line 
of our $SO(10)$ GUT model and their roles.} \label{tab:particle_content}}
\end{table}

The ${\bf 24}$-plet Higgs \footnote{As an adjoint representation, it could be replaced by a Wilson line which appears as a ${\bf 24}$-plet of $SU(5)$. Below the scale of $SU(5)$ breaking, either the Higgs or a Wilson line appears a VEV, and thus, there is no difference when discussing phenomenological issues  below the GUT scale.} in the last row of Table~\ref{tab:particle_content} is decomposed to the subgroup $SU(3)_c\times SU(2)_L \times U(1)_y \times U(1)_\chi $ in the following way,
\begin{eqnarray}
({\bf24},0) &\to& (\overline{\bf3},{\bf2},{\frac 56, 0}) +({\bf3},{\bf2},{-\frac 56, 0}) + ({\bf8},{\bf1},{0,0}) + ({\bf1},{\bf3},{0,0}) + ({\bf1},{\bf1},{0,0}) 
\label{W24}
\end{eqnarray}
It includes a trivial singlet of the subgroup. Thus, a non-vanishing VEV along the direction of $({\bf1},{\bf1},{0,0})$ can break $G^{\rm flip,S}_{51}$ to $G^{\rm S}_{3211}$. This VEV alignment is achieved similarly  to that in the unflipped SUSY $SU(5)$ model, with superpotential shown in \cite{Dimopoulos:1981zb}.

The representations containing the SM matter fields, following the breaking chain $SU(5) \times U(1)_\chi \to SU(3)_c\times SU(2)_L \times U(1)_y \times U(1)_\chi \to  SU(3)_c\times SU(2)_L \times U(1)_Y$ (without considering SUSY), are decomposed as
\begin{eqnarray}
	({\bf10},{-\frac 12}) &\to& ({\bf3},{\bf2},{\frac 16, -\frac 12}) +(\bar{{\bf3}},{\bf1},{-\frac 23, -\frac 12})+({\bf1},{\bf1},{1,-\frac 12}) \,,
	\nonumber\\
&\to& ({\bf3},{\bf2},\frac 16) +(\bar{{\bf3}},{\bf1},\frac 13)+({\bf1},{\bf1},0) = Q+d^c+\nu^c\\
%%%%%%%%%%%%%%%%%%%%%%%%%%%%%
	(\bar{{\bf5}},{+\frac 32}) &\to&	(\bar{{\bf3}}, {\bf1}, \frac 13, \frac 32) + ({\bf1}, {\bf2}, -\frac 12, \frac 32)
 \to
(\bar{{\bf3}}, {\bf1}, -\frac 23) + ({\bf1}, {\bf2}, -\frac 12) = u^c + L\,,
	\nonumber\\
%%%%%%%%%%%%%%%%%%%%%%%%%%%%%
({\bf1},{-\frac 52}) &\to&	({\bf1}, {\bf1}, 0, -\frac 52)
 \to ({\bf1}, {\bf1}, 1) = e^c~,\nonumber
\end{eqnarray}
where $Q=(u,d)$ and $L = (\nu, e)$.
%%%%%%%%%%%%%%%%%%%%%%%%%%%%%
The Higgs representations   are decomposed as follows
\begin{eqnarray}
	({\bf5},+1) &\to& ({\bf3}, {\bf1}, -\frac 13, 1) + ({\bf1}, {\bf2}, \frac 12, 1) \to
({\bf3}, {\bf1}, -\frac 13) + ({\bf1}, {\bf2}, -\frac 12) = D + h_{\rm SM} \label{5higgs}\\
(\overline{\bf10},{+\frac 12}) &\to& (\overline{\bf3},{\bf2},{-\frac 16, \frac 12}) +({\bf3},{\bf1},{\frac 23, \frac 12})+({\bf1},{\bf1},{-1,\frac 12}) \,,
	\nonumber\\
&\to& ({\bf3},{\bf2},\frac 16) +(\bar{{\bf3}},{\bf1},\frac 13)+({\bf1},{\bf1},0) = \bar{Q}_H+\bar{d}^c_H+\bar{\nu}^c_H \label{10higgs}\,,
\end{eqnarray}
where, $h_{\rm SM}$ is the SM EW-doublet Higgs, $D$ is a heavy triplet Higgs, and  $\bar{\nu}^c_H$ 
is the scalar whose VEV  breaks $U(1)_y \times U(1)_\chi$ to $U(1)_Y$ and gives masses to the RH neutrinos.
All chiral fermions are shown in the left-handed convention. In the supersymmetric case, the Higgs sector -in addition to (\ref{5higgs}) and (\ref{10higgs})- includes also the representations $({\bf \bar 5},-1)$ and $({\bf {10}},-1/2)$. In this case, for later convenience we shall use the notation 
$ H\equiv ({\bf {10}}, -1/2),$ and $ \bar{H}\equiv ({\bf \overline{10}}, 1/2)$.
Moreover, for the fermion generations we will adopt the symbols 
$F_i\equiv ({\bf {10}}, -\frac 12)_i, \bar{f}_j= (\bar{{\bf5}},3/2)_j $ where $i,j$ are fermion generation indices.
Notice also that in flipped $SU(5)$ the MSSM Higgs doublets  $h_d$ and $ h_u$  reside in $h=({\bf  5},1)$ and $\bar h=({\bf \bar 5},-1)$  representations respectively. 

The flipped $SU(5)$ is also regarded as a symmetry breaking pattern of the $SO(10)$ gauge group, alternative to the Georgi-Glashow $SU(5)$ or the Pati-Salam chain.
Representations of fields in the group $SU(5)\times U(1)_{\chi}$ are achieved following the decomposition of representations of $SO(10)\to SU(5)\times U(1)_{\chi}$,
\begin{eqnarray}
	\underline{\bf16}&\to& ({\bf10},-\frac 12) +(\bar{{\bf5}},{\frac 32})+({\bf1},{-\frac 52}) \,,
	\nonumber\\
	\underline{\bf10}&\to& ({\bf5},1)+(\overline{{\bf5}},{-1}) 
\,.
\end{eqnarray}

%%%%%%%%%%%%%%%%%%%%%%%%%%%%%%%%%%%%%%%%%%%%%%%%%%%%%%%%%%%%%%%%%

The fermion masses arise from the following $SU(5)\times U(1)_{\chi}$ invariant couplings
\begin{eqnarray}
\label{eq:FermCoupl}
{\cal W}_d&=&(Y_d)^*_{ij}\,({\bf10},{-\frac 12 })_i \cdot ({\bf10},{-\frac 12})_j \cdot ({\bf5},{1})\;\to \; (Y_d)^*_{ij}\, Q_i\,d^c_j\,h_d \,, \nonumber\\
{\cal W}_u&=& (Y_u)^*_{ij}\, ({\bf10},{-\frac 12})_i \cdot (\bar{{\bf5}},{\frac 32})_j \cdot (\bar{{\bf5}},{-1})\;\to \; (Y_u)^*_{ij}\, [ Q_i\,u^c_j +\nu^c_i\,L_j ]\, h_u \,, \nonumber\\
{\cal W}_l&=&(Y_l)^*_{ij}\, ({\bf1},{-\frac 52})_j \cdot (\bar{{\bf5}}, {\frac 32})_i \cdot ({\bf5},1)\;\rightarrow \; (Y_l)^*_{ij}\, e^c_j\,L_i \,h_d \,.
\end{eqnarray}
Here $Y_u$, $Y_d$ and $Y_l$ are $3\times 3$ Yukawa matrices and $Y_d$ is symmetric. These coefficient matrices are introduced with a complex conjugation to match with the SM left-right  non-SUSY convention. 
The Dirac Yukawa coupling matrix $Y_\nu$ is read from the above equation as $Y_\nu = Y_u^T$. 
Also, a higher order term providing Majorana masses  for the right-handed neutrinos can be written
\begin{eqnarray}
{\cal W}_{\nu^c}&=&(\lambda^{\nu^c})^*_{ij}\frac{1}{2M_S}\,{\bar H}\,{\bar H}\,{F_i}\,F_j\to \frac12 (M_{\nu^c})_{ij}^*\nu^c_i\nu^c_j \,.
\label{Maj}
\end{eqnarray}
where $(M_{\nu^c})_{ij} =( \lambda^{\nu^c})_{ij} \frac{\langle \bar\nu^c_H\rangle^2}{M_S}$. The light neutrinos gain masses via the usual type-I seesaw mechanism,
\begin{eqnarray}
M_{\nu} = - Y_\nu M_{\nu^c}^{-1} Y_\nu^T \langle h_u \rangle^2 \,.
\end{eqnarray}
If additional singlet fields $\nu_S, \Phi_i$ are present (which is the usual case in
string derived models), then -depending on their specific properties- the
following couplings could be generated
\begin{eqnarray}
F\bar H \nu_S+ \bar h h \Phi_i + \lambda_{ijk} \Phi_i\Phi_j\Phi_k
+\cdots
\end{eqnarray}
In addition to SM representations, the Higgs sector contains
dangerous colour triplets $D_h, D_H+c.c.$. They become massive
through the following terms
\begin{eqnarray}
HHh +\bar H\bar H\bar h \to  \langle \nu^c_H\rangle  D_H^c D_h
+ \langle \bar \nu^c_H\rangle \bar D_H^c  \bar  D_h \,.
\end{eqnarray}

Note that, if no other symmetry exists the  terms such as
$H F_i h,  H \bar f_j \bar h$ could be possible. Such terms would generate
dangerous mixing between Higgs and Matter fields:
\begin{eqnarray}    (a F+b  H) \bar f_j \, h +\cdots
\end{eqnarray}
Remarkably, a softly broken $Z_2$ symmetry~\cite{Antoniadis:1987dx} which is odd only for the Higgs field $H\to -H$, suppresses all these couplings from the Lagrangian, while all the previous (useful) terms are left intact.\footnote{This $Z_2$ must be softly broken to avoid the domain wall problem after the spontaneous breaking of $Z_2$ once $H$ gains the VEV.}
Similar symmetries have been discussed in~\cite{Kyae:2005nv}~.

For rank-one mass textures the couplings in Eq.~\eqref{eq:FermCoupl} predict $m_t=m_{\nu_{\tau}}$ at the GUT scale.
However, in contrast to the standard $SU(5)$ model, down quark and lepton mass matrices are not
related, since at the $SU(5)\times U(1)_{\chi}$ level they originate from different Yukawa
couplings. This is an important difference with the ordinary $SU(5)$. We know that
in order to obtain the observed lepton and down quark mass spectrum at low energies,
at the GUT scale the following relations should hold \cite{Georgi:1979df}
\begin{eqnarray}
m_{\tau}=m_b\,,\quad
m_{\mu}=3\,m_s\,.
\end{eqnarray}
In the ordinary $SU(5)$, the masses are related and the relations
can be attributed to the Higgs adjoint which couples differently.  This mechanism though
is not operative in the minimal flipped $SU(5)$ due to the absence of the adjoint, as noticed above, thus
the mass matrices are not related and Yukawas should be adjusted accordingly.

We further review the derivation of the matching condition between gauge couplings of $U(1)_Y$ and those of $SU(5) \times U(1)_\chi$. Computing the traces and finding normalisation constants so that the 
final trace is 2, give:
\begin{eqnarray}
C_y^2y^2&=&C_y^2 \frac{10}3 =2 \to C_y=\sqrt{\frac 35} \,,
\nonumber\\
C_{\chi}^2\chi^2&=&C_\chi^2 20 =2 \to C_{\chi}=\frac{1}{\sqrt{10}} \,\cdot 
\end{eqnarray}
In terms of normalised  generators, the hypercharge is written as 
 $Y= \frac{1}{5C_y}\left( \tilde y+\kappa \tilde\chi\right)$, 
where the ratio is
$\kappa\equiv 2 \frac{C_{y}}{C_{\chi}} = 2\sqrt{6} $.
Finally  $\tilde{Y}=\sqrt{\frac 35}Y$ implies
\begin{eqnarray}
\tilde Y= \frac 15 \left( \tilde y+2 \sqrt{6}\, \tilde\chi\right)~,
\end{eqnarray}
and for the $U(1)_Y$  gauge coupling
\begin{eqnarray}
(1+\kappa^2) \frac{1}{a_Y}= \frac{1}{a_5}+\kappa^2 \frac{1}{a_{\chi}}
\end{eqnarray}
or equivalently,
\begin{eqnarray}
\frac{1}{\alpha_{\tilde Y}}=\frac{1}{25}\frac{1}{\alpha_{\tilde y}}+\frac{24}{25} \frac{1}{\alpha_{\tilde \chi}}\,.
\end{eqnarray}
For initial values $a_{\chi}=a_5$, we obtain the standard relation
of $SU(5)$.   In general, however, $a_{\chi}\ne a_5$, and there is more flexibility.

\section{Two-loop RG running}\label{sec:gauge_unifcation}

Renormalisation group (RG) running of the gauge couplings can be briefly described in the follow picture. As the energy scale $Q$ runs from one intermediate symmetry breaking scale to its neighbouring intermediate symmetry breaking scale (e.g., from $M_{B-L}$ to $M_{\rm GUT}$), the gauge coupling varies continuously along with $Q$. For energy scale jumps across the intermediate symmetry breaking scale (e.g., $M_{B-L}$), gauge couplings below  and above the scale satisfy matching conditions.

Given a gauge symmetry $G \equiv H_1 \times \cdots \times H_n$ as a product of simple Lie groups $H_i$ for $i=1,2, \cdots$ between two neighbouring symmetry breaking scales $Q_1$ and $Q_2$, the two-loop RG equation of the gauge coupling $g_i$ for the group $H_i$ is given by
\begin{eqnarray}
\frac{d\alpha_i}{d t} = \beta_i (\alpha_i)\,,
\end{eqnarray}
where $\alpha_i =g_i^2/(4\pi)$, $t = \log (Q/Q_0)$ is the logarithm of the energy scale $Q$ for $Q_1 < Q <Q_2$, and $Q_0$ is an arbitrary energy scale here fixed at the $Z$ pole mass. The $\beta$ function on the right hand side is written as
\begin{eqnarray}
\beta_i = - \frac{1}{2\pi} \alpha_i^2 ( b_i + \frac{1}{4\pi} \sum_{j} b_{ij} \alpha_j )~,
\end{eqnarray}
up to the two-loop level. Here, $b_i$ and $b_{ij}$ are the normalised coefficients of one- and two-loop contributions of fields (or superfields), respectively. Both are determined by the group $H_i$ and representations  of particles in $H_i$, and $b_{ij}$ further depend  on the particle representation in $H_j$. In this work, we will simply list their values when they are used for given particle contents in our model, which are given in Table~\ref{tab:decomposition}. For each energy-scale interval from the EW scale to the scale of flipped $SU(5)$ breaking in this table, the beta coefficients $b_i$ and $b_{ij}$ are given in Table~\ref{tab:beta}. These coefficients are calculated in the following way. For example, for RG running from $G_{\rm MSSM}$ down to $G_{\rm SM}$, $b_i$ and $b_{ij}$ are calculated by including only fields preserving $G_{\rm SM}$ in Table~\ref{tab:decomposition}. We refer to \cite{King:2021gmj} for more general discussions on how to calculate these coefficients.
The above differential equation has analytical solution
\begin{eqnarray}
\alpha_i^{-1}(t) = \alpha_i^{-1}(t_1) - \frac{b_i}{2\pi} (t-t_1) + \sum_j \frac{b_{ij}}{4\pi b_i} \log\left(1- \frac{b_j}{2\pi} \alpha_j(t_1) \, (t-t_1) \right) \,,
\end{eqnarray}
if the condition $b_j \alpha_j |t_2 - t_1| <1$ is satisfied and no particle decouples for $t$ varying  from $t_1$ to $t_2$~\cite{Bertolini:2009qj}.

\begin{table}[t!].
\begin{center}
\begin{tabular}{| c | c | c |}
\hline \hline
Model & Matter (Super-)Fields & Higgs (Super-)Fields
\\\hline & & \\[-4mm]
$G_{51}^{\rm flip,S}$ &
$({\bf 10}, - \frac12)+ (\overline{\bf 5}, + \frac{3}{2})+ ({\bf 1}, - \frac52)$ & $({\bf 5}, +1)+  (\overline{\bf 10}, + \frac12)$ \\
& & \\[-4mm]\hline
\multirow{2}*{$G^{\rm S}_{3211}$} &
$({\bf3},{\bf2},{\frac 16, -\frac 12}) +(\bar{{\bf3}},{\bf1},{-\frac 23, -\frac 12})+({\bf1},{\bf1},{1,-\frac 12})$ & $({\bf1}, {\bf2}, \frac 12, 1)
+ ({\bf1}, {\bf2}, -\frac 12, -1)$ \\
& $+ (\bar{{\bf3}}, {\bf1}, \frac 13, \frac 32) + ({\bf1}, {\bf2}, -\frac 12, \frac 32) +({\bf1}, {\bf1}, 0, -\frac 52)$ & $+({\bf1},{\bf1},{-1,\frac 12})+({\bf1},{\bf1},{1,-\frac 12})$ \\[1mm]\hline
& \\[-5mm]
\multirow{2}*{$G_{\rm MSSM}$} &
$({\bf3},{\bf2},\frac 16) +(\bar{{\bf3}},{\bf1},\frac 13)+({\bf1},{\bf1},0)$ & $({\bf1}, {\bf2}, -\frac 12)+({\bf1}, {\bf2}, \frac 12)$ \\
& $(\bar{{\bf3}}, {\bf1}, -\frac 23) + ({\bf1}, {\bf2}, -\frac 12) + ({\bf1}, {\bf1}, 1)$ &
\\[1mm]\hline
& \\[-5mm]
\multirow{2}*{$G_{\rm SM}$} &
$({\bf3},{\bf2},\frac 16) +(\bar{{\bf3}},{\bf1},\frac 13)+({\bf1},{\bf1},0)$ & $({\bf1}, {\bf2}, \frac 12)$ \\
& $(\bar{{\bf3}}, {\bf1}, -\frac 23) + ({\bf1}, {\bf2}, -\frac 12) + ({\bf1}, {\bf1}, 1)$ &
\\[1mm]
\hline \hline
\end{tabular}
\end{center}
\caption{{Decomposition of the matter multiplet ${\bf 16}$ and Higgses in each step of the breaking chain. Note that the adjoint ${\bf 24}$ Wilson line is not included in the table as it is not a field.} \label{tab:decomposition}}
\end{table}

\begin{table}[tbp]
\begin{center}
\begin{tabular}{| c l |}
\hline \hline
$G_{51}^{\rm flip,S}$ & broken at $Q=M_{\rm GUT}$ \\\hline
& \\[-5mm]
$\Bigg\downarrow$ &
$\{b_i\} = \begin{pmatrix} -3 \\ 1 \\ \frac{39}{5} \\ \frac{129}{20} \end{pmatrix} \,, \quad
\{b_{ij} \} = \begin{pmatrix}
 14 & 9 & \frac{11}{5} & \frac{9}{5} \\
 24 & 25 & \frac{9}{5} & \frac{11}{5} \\
 \frac{88}{5} & \frac{27}{5} & \frac{271}{25} & \frac{54}{25} \\
 \frac{72}{5} & \frac{33}{5} & \frac{54}{25} & \frac{1593}{200} \\
\end{pmatrix}$ \\[-5mm] &
\\\hline
$G^{\rm S}_{3211}$ & broken at $Q=M_{B-L}$ \\\hline
& \\[-5mm]
$\Bigg\downarrow$ &
$\{b_i\} = \begin{pmatrix} -3 \\ 1 \\ \frac{33}{5} \end{pmatrix} \,,\quad
\{b_{ij} \} = \begin{pmatrix}
 14 & 9 & \frac{11}{5} \\
 24 & 25 & \frac{9}{5} \\
 \frac{88}{5} & \frac{27}{5} & \frac{199}{25} \\
\end{pmatrix}$
\\[-5mm] & \\\hline
$G_{\rm MSSM}$ & broken at $Q=M_{\rm SUSY}$ \\\hline
& \\[-5mm]
$\Bigg\downarrow$ &
$\{b_i\} = \begin{pmatrix} -7 \\ -\frac{19}{6} \\ \frac{41}{10} \end{pmatrix} \,,\quad
\{b_{ij} \} = \begin{pmatrix}
-26 & \frac{9}{2} & \frac{11}{10} \\
12 & \frac{35}{6} & \frac{9}{10} \\
\frac{44}{5} & \frac{17}{10} & \frac{199}{50}
\end{pmatrix}$
\\[-5mm] & \\\hline
$G_{\rm SM}$ & \\
\hline \hline
\end{tabular}
\end{center}
\caption{Coefficients $b_i$ and $b_{ij}$ of gauge coupling $\beta$ functions appearing in the specified breaking chain.  \label{tab:beta}}
\end{table}

During a symmetry breaking at a certain scale (e.g., $M_{\rm GUT}$, $M_{B-L}$ or $M_{\rm SUSY}$), the gauge couplings of the larger symmetry and those of the residual symmetry after the symmetry breaking satisfy matching conditions. Below we list one-loop matching conditions which appear in the GUT breaking chains. In this work, we are encountering two types of symmetry breaking: 1) at $M_{B-L}$ and $M_{\rm GUT}$, the gauge symmetry breaking, where $M_{B-L}$ and $M_{\rm GUT}$ are identified as corresponding gauge boson masses; 2)at $M_{\rm SUSY}$, the SUSY breaking, where we assume a unified mass scale for all superpartners and this scale is recognised as $M_{\rm SUSY}$.\footnote{A unified mass scale for all SUSY particles at $M_{\rm SUSY}$ should be considered as a rough assumption here. In SUSY GUTs, a dark matter candidate is identified with the lightest SUSY particle provided  it is  neutral~\cite{Low:2014cba,Fukuda:2018ufg}. Restriction on the dark matter overproduction sets an upper bound around 10~TeV on the dark matter mass \cite{Profumo:2005xd}, and also on the SUSY scale. By splitting the mass spectrum of SUSY particles, e.g., that in the split SUSY \cite{Giudice:2004tc}, the SUSY scale is relaxed and in general can be treated as a free parameter varying up to the GUT scale. Note that splitting the mass spectrum slightly modifies the $\beta$ coefficients $b_i$ and $b_{ij}$ at scale lower than $M_{\rm SUSY}$; seeing e.g., our recent discussion in \cite{Fu:2023mdu}. Its influence on the RG running behaviour of the gauge couplings is small and goes beyond the main point of this paper. }
Matching conditions at the scale of gauge symmetry breaking are given as follows. For a simple Lie group $SU(m)$ broken to subgroup $SU(n)$ ($m>n>1$) at the scale $Q = M_I$, the one-loop matching condition is given by \cite{Chakrabortty:2017mgi}
\begin{eqnarray}
SU(m) \to SU(n)\,: \quad
\alpha_{SU(m)}^{-1}(M_I) - \frac{m}{12\pi} &=& \alpha_{SU(n)}^{-1}(M_I) - \frac{n}{12\pi} \,,
\end{eqnarray}
where the quadratic Casimir for the adjoint presentation of the $SU(n)$ group, $C_2(SU(n)) = n$, has been used.
For $G^{\rm S}_{3211}\to G_{\rm SM}$, we encounter the breaking, $U(1)_y \times U(1)_\chi \to U(1)_Y$, which has the matching condition:
\begin{eqnarray} \label{eq:matching_U1}
\frac{1}{25}\alpha_{\tilde y}^{-1}(M_{B-L})  + \frac{24}{25}\alpha_{\tilde \chi}^{-1}(M_{B-L}) = \alpha_{\tilde Y}^{-1}(M_{B-L}) \,.
\end{eqnarray}
The matching condition at the SUSY breaking scale $M_{\rm SUSY}$ is induced by the usage of different renormalisation schemes for the scale below and above $M_{\rm SUSY}$. At $Q< M_{\rm SUSY}$, the RGE was derived in the usual $\overline{\rm MS}$ scheme. At $Q> M_{\rm SUSY}$, it was calculated in the $\overline{\rm DR}$ scheme, where SUSY can be preserved \cite{Martin:1993yx}. The relation of couplings in the two schemes is described by
\begin{eqnarray}
\alpha_{\rm DR}^{-1} &=& \alpha_{\rm MS}^{-1} - \frac{1}{12\pi} C_2(H_i) \,,
\end{eqnarray}
where $H_i$ is a Lie group. For $H_i = U(1)$, $C_2(U(1))=0$, whilst  $C_2(SU(n))$ has been given above.

At the EW scale, we set the SM gauge couplings at their best-fit values, i.e.,
$\alpha_3 = 0.1184$, $\alpha_2 = 0.033819$ and $\alpha_1 = 0.010168$ \cite{Xing:2011aa}. Applying the matching conditions at intermediate scales and running RGEs in the interval between each two scales nearby, gauge couplings can run from the EW scale to a sufficient high scale. 

In the flipped $SU(5)$ framework, gauge couplings of $SU(3)_c$, $SU(2)_L$ and $U(1)_y$ are supposed to be unified into a single gauge coupling, i.e., $g_5$ with $\alpha_5 \equiv g_5^2/{4\pi}$ satisfying
\begin{eqnarray} \label{eq:unification}
\alpha^{-1}_3(M_{\rm GUT}) - \frac{3}{12\pi} = \alpha^{-1}_2(M_{\rm GUT}) - \frac{2}{12\pi} = \alpha^{-1}_{\tilde y}(M_{\rm GUT})
\end{eqnarray} 
Some interesting features of correlations of $M_{\rm SUSY}$, $M_{B-L}$ and $M_{\rm GUT}$ and the gauge couplings $g_y$ and $g_\chi$ can be summarised below.
\begin{itemize}
\item
As seen in table~\ref{tab:beta}, the $\beta$-coefficients $b_i$ and $b_{ij}$ for $SU(3)_c \times SU(2)_L$ do not change for the scale passing $M_{B-L}$. Thus $M_{B-L}$ does not influence the unification of $SU(3)_c \times SU(2)_L \subset SU(5)$ at $M_{\rm GUT}$, and thus $M_{\rm GUT}$ is only correlated with $M_{\rm SUSY}$.
\item 
$g_y$ is correlated with $M_{B-L}$ and $M_{\rm GUT}$ via the second identity of Eq.~\eqref{eq:unification}, and accordingly, $g_\chi$ is correlated with these scales via Eq.~\eqref{eq:matching_U1}. It is also convenient to obtain correlation between $g_{B-L}$ and these scale via the matching condition
\begin{eqnarray} \label{eq:alpha_BL}
\alpha_{B-L}^{-1} = \frac83 \left( \frac25\alpha^{-1}_{\tilde y} + \frac35 \alpha^{-1}_{\tilde \chi} \right)\,,
\end{eqnarray}
where $\alpha_{B-L} = g_{B-L}^2/(4\pi)$ is the usual unnormalised coupling as the factor $8/3$ in the right hand side appears. 
\end{itemize}
Eventually, the unification leaves only two free parameters among scales and gauge couplings. One can choose, e.g., $M_{B-L}$ and $M_{\rm GUT}$, as variables to determine  the rest of the parameters. 

\begin{figure}[t!]
\centering
\includegraphics[height=.45\textwidth]{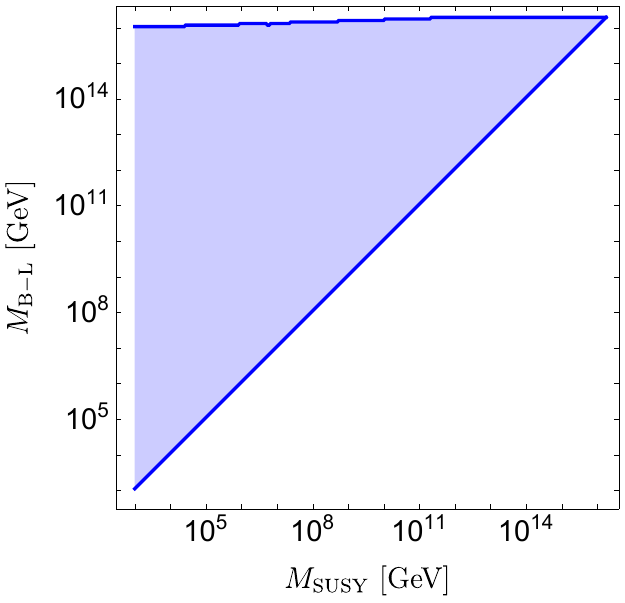}
\includegraphics[height=.45\textwidth]{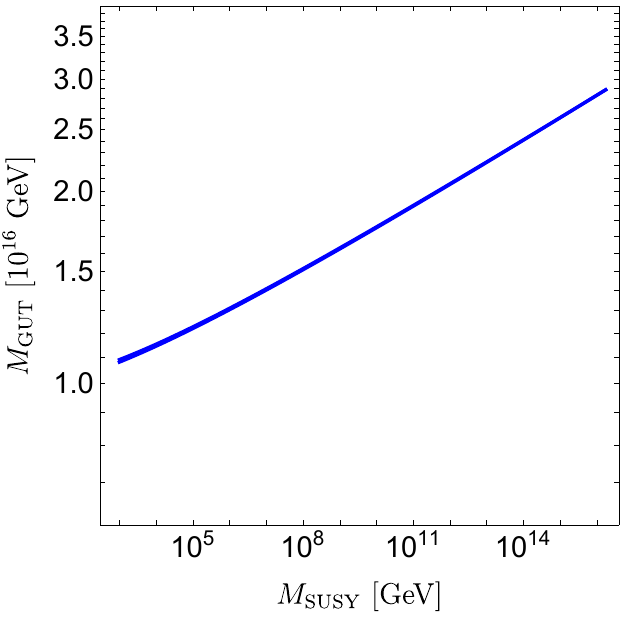}
\caption{\label{fig:predictions} Mass scales restricted by gauge unification in SUSY $SU(5)$ with $1~\text{TeV} < M_{\rm SUSY} < M_{B-L} < M_{\rm GUT}$. }
\end{figure}

\begin{figure}[t!]
\centering
\includegraphics[width=.45\textwidth]{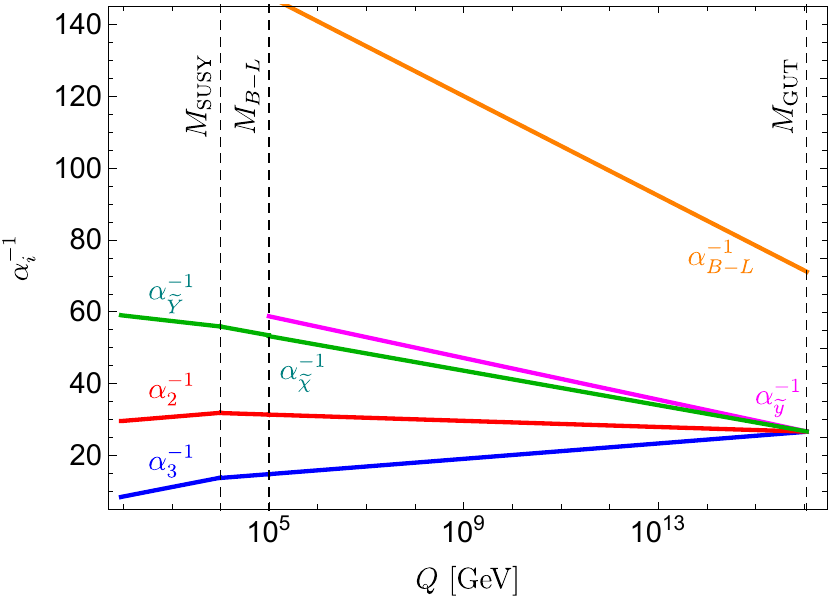}
\includegraphics[width=.45\textwidth]{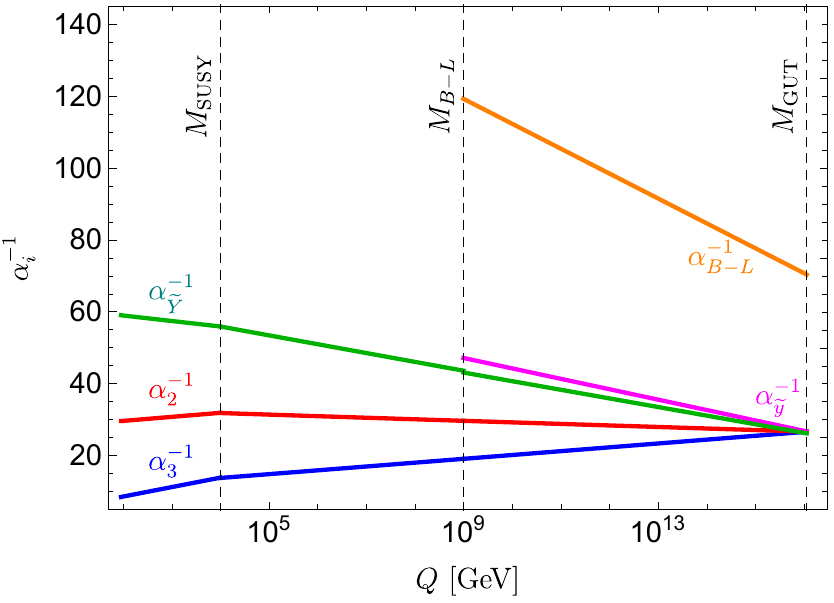}
\includegraphics[width=.45\textwidth]{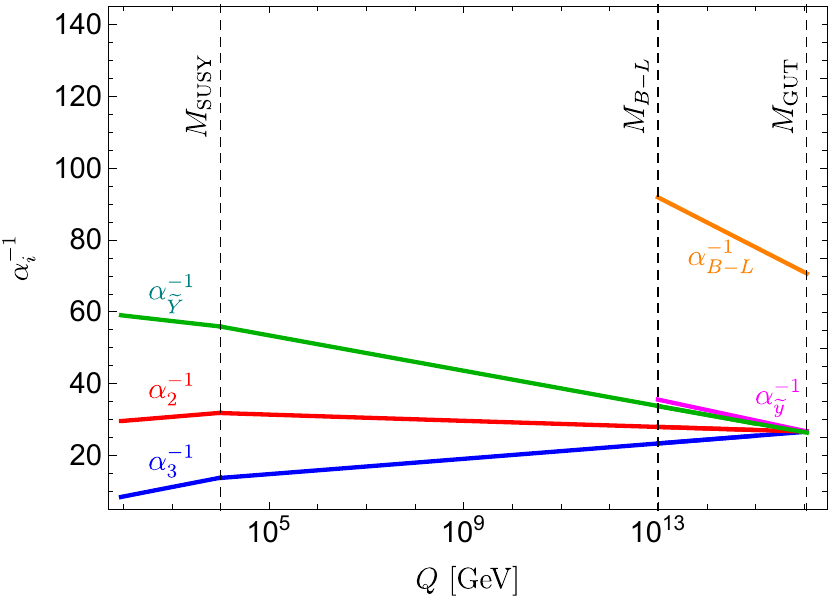}
\includegraphics[width=.45\textwidth]{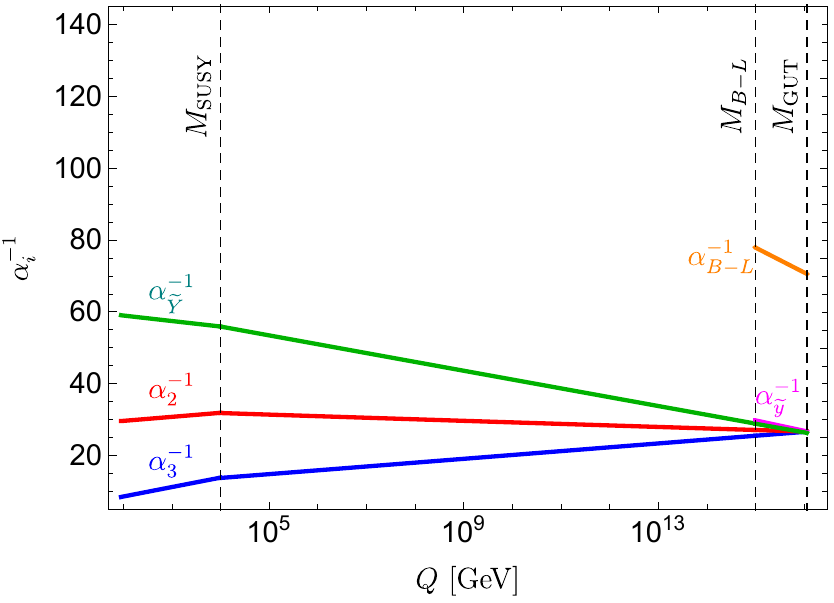}
\caption{\label{fig:RGE} Gauge couplings running with the scale. From the top left to the bottom right, $M_{B-L} = 10^{5}, 10^{9}, 10^{13}, 10^{15}$~GeV are used and $M_{\rm SUSY} = 10$~TeV is fixed. Note that these values are used only for illustration purposes as some of them are not consistent with neutrino mass as will be discussed in section~\ref{sec:flavour}. The $SU(5)$ unification scale  $M_{\rm GUT}$ is solved via the unification of $\alpha_2$ and $\alpha_3$. In the range $M_{B-L} < Q < M_{\rm GUT}$, we show the running of the unnormalised coupling $\alpha_{B-L}$ as the $U(1)_{B-L}$ symmetry is essential to neutrino mass and the generation of cosmic strings.} 
\end{figure}

The above RG running procedure involves three energy scales $M_{\rm MSSM}$, $M_{B-L}$ and $M_{\rm GUT}$. Gauge unification requires that two SM gauge couplings $g_3$ and $g_2$ meet at $M_{\rm GUT}$, which is different from the situation in usual non-flipped $SU(5)$ or $SO(10)$ models that all three SM gauge couplings meet at the GUT scale, up to matching conditions. Furthermore, note that the inclusion of the intermediate scale $M_{B-L}$ does not change the beta coefficients in the $SU(3)_c\times SU(2)_L$ block, as seen in Table~\ref{tab:beta}. As a consequence, $M_{B-L}$ has little influence on the RG running of $g_3$ and $g_2$, and thus is not correlated to the unification scale $M_{\rm GUT}$ except the requirement $M_{\rm GUT} \geqslant M_{B-L}$. 
The allowed parameter space for three energy scales are shown in Fig.~\ref{fig:predictions}, where $1~\text{TeV} \leqslant M_{\rm SUSY} \leqslant M_{B-L} \leqslant M_{\rm GUT}$ is assumed. 

Cases of gauge coupling RG running with the energy scale is shown in Fig~\ref{fig:RGE}, where $M_{\rm SUSY} =10$~TeV is fixed and $M_{B-L}=10^{4}, 10^{9}, 10^{13}, 10^{15}$~GeV are used from the top left to the bottom right panels for illustration. The GUT scale $M_{\rm GUT}$, referring to the breaking of $SU(5)$, is determined by the unification of $\alpha_2$ and $\alpha_3$.  It is seen that varying the scale $M_{B-L}$ does not influence the unification scale $M_{\rm GUT}$. We also show the RG running of the coupling $\alpha_{B-L}$ between $M_{B-L}$ and $M_{\rm GUT}$. Setting $M_{B-L} =10^{15}$~GeV leads to the gauge coupling $g_{B-L} = \sqrt{4\pi/\alpha_{B-L}^{-1}} = 0.40$. Assuming a lower $B-L$ breaking scale could predict a slightly smaller $g_{B-L}$ value. For example, $M_{B-L} =10^{5}$~GeV leads to $g_{B-L} = 0.29$. 

\section{Proton decay}

We discuss the proton decay in the SUSY flipped $SU(5)$ model. Heavy gauge bosons are integrated out at low energy and leave dimension-6 operators. These 4-Fermi interactions violate the baryon number and lead to the decay of nucleons. A typical channel is $p \to \pi^0 e^+$. In the SUSY extension, dimension-5 between two fermions and two superpartners, mediated by colour-triplet Higgs, is another source of baryon number violation. These operators get dressed via loop corrections and might enhance the decay width of another channel $p \to K^+ \bar{\nu}$ greatly. In SUSY $SU(5)$ or $SO(10)$ models, a null result for the observation of proton decay raises the SUSY breaking scale (see discussions e.g. in~\cite{Murayama:2001ur} and \cite{Goh:2003nv,Severson:2015dta}). Furthermore, careful treatment should be paid to avoid the overproduction of the dark matter, which is predicted to be  the lightest supersymmetric particle if it is neutral \cite{Fu:2023mdu}. In SUSY flipped $SU(5)$ models, as will be examined in the coming subsection, the dimension-5 contribution is suppressed due to the missing partner mechanism, thus there is no restriction on the SUSY breaking scale from proton decay. Only dimension-6 contribution need to be considered, as will be done  in section~\ref{sec:D6}. 

\subsection{Dimension-5 contributions} \label{sec:D5}

Due to the missing partner mechanism in Flipped $SU(5)$  the dimension-5 operators are suppressed and as such, they are neglected.
This suppression mechanism works as follows. The extra down-type triplets, arise from the decomposition of the various Higgs representations, and in our particular assignments we have
$$d^c_H\in H,\;  \overline{d^c}_H\in {\bar H},\;  D \in h,\; \bar D\in {\bar h}~.$$

When 	$H+\overline{H}$ acquire vevs,
the tree-level superpotential terms related to the triplet masses  
yield the following mass terms for the triplets
\be \lambda_1 H H h +\lambda_2 \bar H \bar H \bar h \to \lambda_1\langle \nu_H^c\rangle  d^c_H  D + \lambda_2 \langle \bar\nu_H^c\rangle  \overline{d^c}_H \bar D~. \label{Tripletmasses}
 \ee
 	It can readily be observed that due to the symmetries, there are no superpotential terms which couple directly the two triplets 
 	residing in  ${\bf 10}_H, \overline{\bf 10}_{\bar H}$ representations. It is possible however, that suppressed contributions $\sim \frac{m^{n+1}}{M^n_{P}}\bar H H$ might arise at higher (NR)-order.
 	Furthermore, notice that a  superpotential term which couples the two Higgs fiveplets $\mu \bar {\bf 5}_{\bar h} {\bf 5}_h$ implies a direct mass $\mu  \bar D D$ which is suppressed compared to the GUT mass terms~(\ref{Tripletmasses}), i.e., $\mu\ll  \langle \nu_H^c\rangle$. 
 	Assuming for simplicity  $\lambda_1\langle \nu_H^c\rangle=\lambda_2 \langle \bar\nu_H^c\rangle \equiv M_{d_H^cD} $, in the basis $(D,d^c_H)$ we can write the triplet mass matrix as follows 
 	\be 
 	M_{T}=\left(\begin{array}{cc}
 		\mu &m_{d_h^c D}\\
 	m_{d_h^c D}&m_{\bar DD} 
 	\end{array}
 	\right)\;\Rightarrow\;
 	\left(\begin{array}{cc}
 		 a&b\\
 		 b&d 
 		 \end{array}
 	 \right)~,
 	\ee 
 	in a self-explanatory notation. Of course $b> d>a$.
 	We  assume for simplicity that all entries are real, so that we can diagonalise the mass matrix by 
 	\[ V=	\left(\begin{array}{cc}
 		\cos\theta &\sin\theta \\
 		-\sin\theta &\cos\theta 
 	\end{array}
 	\right)~.\]
 Now, a tree-level diagram leading to d-5 oparators mediated by 
 $D$-$\bar D$ implies an effective mass in the propagator
 \be 
\frac{1}{M^2_{\rm eff}}= \sum_{j=1}^2 V_{1j}\frac{1}{M^2_{T_j}}V_{j2}^{\dagger}= 
\frac{(a+d)b}{(ad-b^2)^2} \equiv  \frac{(a+d)b}{({\rm Det}M_T)^2}
\sim \frac{{m_{\bar DD}}}{{m^3_{d^c_HD}}}~.
 \ee 
For the hierarchies given above,  $M_{\rm eff} \gg m_{d^c_HD}> {m_{\bar DD}}$ which implies sufficient suppression to dimension-5 operators.

\subsection{Dimension-6 contributions} \label{sec:D6}

We are then left with dimension-6 operator contributions mediated by heavy gauge bosons. A model-independent description of dimension-6 BNV operators are given by 
\begin{eqnarray} \label{eq:BNV}
&& \frac{1}{\Lambda_1^2} \left[
(\overline{u_R^{c}} \gamma^\mu Q)(\overline{d_R^{c}} \gamma_\mu L) +
(\overline{u_R^{c}} \gamma^\mu Q)(\overline{e_R^{c}} \gamma_\mu Q) \right ] \nonumber\\
&+& \frac{1}{\Lambda_2^2} \left[
(\overline{d_R^{c}} \gamma^\mu Q)(\overline{u_R^{c}} \gamma_\mu L) +
(\overline{d_R^{c}} \gamma^\mu Q)(\overline{\nu_R^{c}} \gamma_\mu Q) \right ]\,.
\end{eqnarray}
$\Lambda_1$ and $\Lambda_2$ are heavy mass scales determined by the GUT model. In the usual non-flipped $SU(5)$, $\Lambda_1 = M_{\rm GUT}/g_{\rm GUT}$ and $\Lambda_2 \to \infty$; In flipped $SU(5)$, $\Lambda_2 = M_{\rm GUT}/g_{\rm GUT}$ and $\Lambda_1 \to \infty$;  In $SO(10)$, $\Lambda_1 = \Lambda_2 = M_{\rm GUT} /g_{\rm GUT}$, where $g_{\rm GUT}$ and $M_{\rm GUT}$ should be understood as gauge couplings and heavy gauge boson  masses of the relevant gauge group. 

We consider dimension-6 operators contributing to the typical channel $p \to \pi^0 e^+$ in flipped $SU(5)$. As seen in Fig.~\ref{fig:predictions}, the lower bound of $M_{\rm GUT}$, due to the requirement of gauge unification, is always larger than $10^{16}$~GeV, which is high enough to avoid current known experimental constraints set by Super-K, $\tau(p \to \pi^0 e^+) > 2.4 \times 10^{34}~{\rm yr}$ \cite{Super-Kamiokande:2020wjk}. 
A further suppression in the flipped $SU(5)$ is the absence of the decay to  a right-handed charged lepton. Since $e_R$ is arranged as a gauge singlet in the model, there is no gauge boson  mediating BNV interactions to $e_R$. This is seen in the second row of  Eq.~\eqref{eq:BNV}, where only the left-handed $e_L$ is involved in the operator. As a comparison, in the unflipped $SU(5)$ or $SO(10)$, the two operators in the first row of Eq.~\eqref{eq:BNV} include the charged lepton, either $e_L$ or $e_R$. Thus, they both contribute to the decay $p \to \pi^0 e^+$. 

The lifetime referring to this channel is defined as $\tau(p\to \pi^0 e^+) = 1/\Gamma(p\to \pi^0 e^+)$ with the partial decay width 
\begin{eqnarray}
\Gamma(p\to \pi^0 e^+) = \frac{m_p}{32\pi} \left(1-\frac{m_\pi^2}{m_p^2}\right)^2 \frac{g_5^4}{M_{\rm GUT}^4} A_L^2 A_{S_1}^2( \langle \pi^0|(ud)_R u_L|p\rangle_e )^2~ |V_{ud}|^2 |(U_l)_{11}|^2~,
\end{eqnarray}
where $A_L = 1.247$ is the long range effect, and the short range effect
\begin{eqnarray}
A_{S_1} &=& \left[ \frac{\alpha_3(M_{\rm SUSY})}{\alpha_3(M_{\rm GUT})} \right]^{\frac{4}{9}}
\left[ \frac{\alpha_2(M_{\rm SUSY})}{\alpha_2(M_{\rm GUT})} \right]^{-\frac{3}{2}}
\left[ \frac{\alpha_{\tilde Y}(M_{\rm SUSY})}{\alpha_{\tilde Y}(M_{\rm GUT})} \right]^{\frac{1}{18}} \nonumber\\
&\times& \left[ \frac{\alpha_3(m_Z)}{\alpha_3(M_{\rm SUSY})} \right]^{\frac{2}{7}}
\left[ \frac{\alpha_2(m_Z)}{\alpha_2(M_{\rm SUSY})} \right]^{-\frac{27}{38}}
\left[ \frac{\alpha_{\tilde Y}(m_Z)}{\alpha_{\tilde Y}(M_{\rm SUSY})} \right]^{-\frac{11}{82}}~,
\end{eqnarray}
depending on energy scales. $\langle \pi^0|(ud)_R u_L|p\rangle_e = -0.131$ is the relevant hadronic matrix element appearing in the channel $p \to \pi^0 e^+$, derived via lattice simulation \cite{Aoki:2017puj}.

\begin{figure}[t!]
\centering
\includegraphics[width=.8\textwidth]{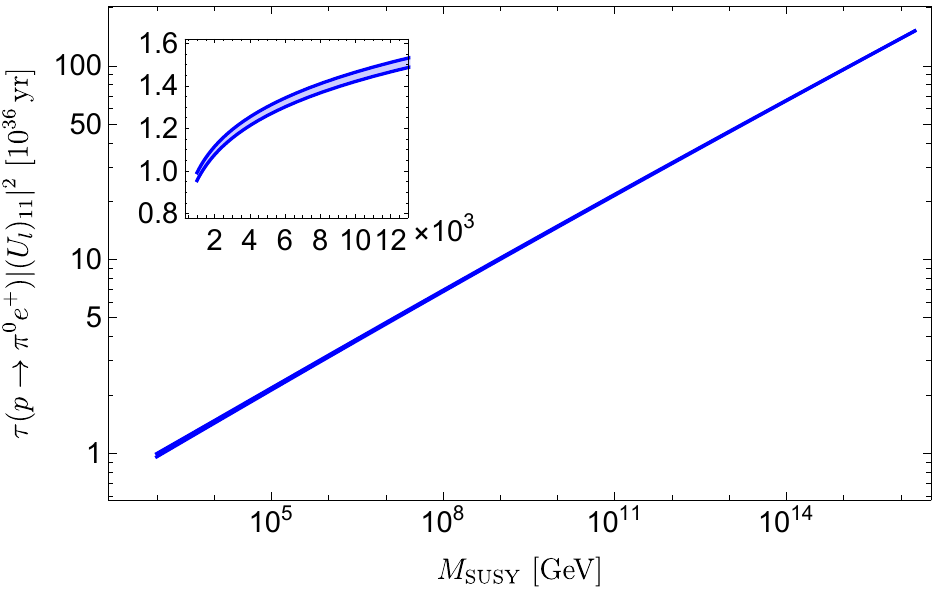}
\caption{\label{fig:proton_decay} Prediction of partial proton lifetime via the channel $p 
\to \pi^0 e^+$  vs the SUSY scale. }
\end{figure}

Taking the gauge unification in the last section into account, we have scanned all mass scales and confirmed that $\Gamma(p\to \pi^0 e^+)/|(U_l)_{11}|^2$ is no more than $[10^{36}~{\rm yr}]^{-1}$, seeing in Fig.~\ref{fig:proton_decay}. This decay width is in general small enough and cannot be excluded by the future proton decay measurement including Hyper-K, unless $|(U_l)_{11}|^2 <0.1$. 

\begin{figure}[t!]
\centering
\includegraphics[width=.8\textwidth]{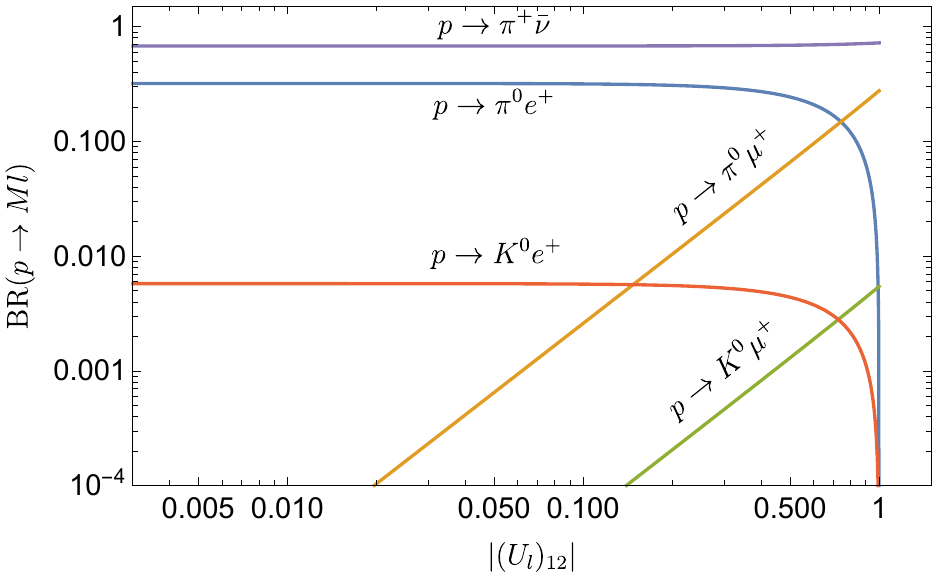}
\caption{\label{fig:proton_decay_BR} Branching ratios for different proton decay channels as a function of $|(U_l)_{12}|$. }
\end{figure}

Following the formulation in \cite{Ellis:2020qad}, we list ratios of all other proton decay channels to $p \to \pi^0 e^+$ as 
\begin{eqnarray} \label{eq:decay_ratio}
\frac{\Gamma(p \to \pi^0 \mu^+)}{\Gamma(p \to \pi^0 e^+)}&=&\left[\frac{\langle \pi^0 | (ud)_R u_L |p\rangle_\mu}{\langle \pi^0| (ud)_R u_L | p \rangle_e}\right]^2\frac{|(U_l)_{12}|^2 }{|(U_l)_{11}|^2} \simeq 0.811 \frac{|(U_l)_{12}|^2 }{|(U_l)_{11}|^2} \,, \nonumber\\
\frac{\Gamma(p \to K^0 e^+)}{\Gamma(p \to \pi^0 e^+)}&=&\left[\frac{\langle K^0 | (us)_R u_L |p\rangle_e}{\langle \pi^0| (ud)_R u_L | p \rangle_e}\right]^2\frac{|V_{us}|^2}{|V_{ud}|^2} \simeq 0.018 \,, \nonumber\\
\frac{\Gamma(p \to K^0 \mu^+)}{\Gamma(p \to \pi^0 e^+)}&=&\left[\frac{\langle K^0 | (us)_R u_L |p\rangle_\mu}{\langle \pi^0| (ud)_R u_L | p \rangle_e}\right]^2\frac{|V_{us}|^2|(U_l)_{12}|^2 }{|V_{ud}|^2|(U_l)_{11}|^2} \simeq 0.016 \frac{|(U_l)_{12}|^2 }{|(U_l)_{11}|^2} \,, \nonumber\\
\frac{\Gamma(p \to \pi^+ \bar{\nu})}{\Gamma(p \to \pi^0 e^+)}&=&\left[\frac{\langle \pi^+ | (ud)_R d_L |p\rangle}{\langle \pi^0| (ud)_R u_L | p \rangle_e}\right]^2\frac{1}{|V_{ud}|^2 |(U_l)_{11}|^2} \simeq \frac{2.12}{|(U_l)_{11}|^2}~\cdot
\end{eqnarray}
Here, $\Gamma(p \to \pi^+ \bar{\nu})$ represents $\sum_i \Gamma(p \to \pi^+ \bar{\nu}_i)$. 
$p \to K^+\bar{\nu}_i$ is not listed as it is proven to be forbidden by the unitarity of the CKM matrix \cite{Ellis:2020qad}. On the right-hand side of the equation, all nuclear matrix element values are taken their numerical values obtained from the lattice simulation \cite{Aoki:2017puj}. 
As a consequence, all ratios depend on the undetermined unitary matrix $U_l$. It is convenient to write out the branching ratio ${\rm BR}(p \to M\, l) = \Gamma(p \to M\, l)/ \Gamma_p$, where $M$ is a meson and $l$ is a lepton, and $\Gamma_p$ is the total decay width of the proton. Considering a natural scenario $|(U_l)_{13}| \ll |(U_l)_{12}|$ and thus, $|(U_l)_{11}| \simeq \sqrt{1-|(U_l)_{12}|^2}$. All BRs then depend  only on the off-diagonal entry $|(U_l)_{12}|$. We show BRs for all channels as functions of $|(U_l)_{12}|$ in Fig.~\ref{fig:proton_decay_BR}.
It is a special feature in flipped $SU(5)$ that the BR of the anti-neutrino channel, $p\to \pi^+ \bar{\nu}$ is larger than the positron channel, as both $\nu_R$ and $\nu_L$ are involved in BNV operators. Varying $|(U_l)_{12}|$ in $(0,1)$, we obtain the branching of this neutrino channel around $70\%$. 

To end the section, we mention the neutron BNV decay. The dominant channel is $n \to \pi^0 \bar{\nu}$ with decay width $\Gamma(n \to \pi^0 \bar{\nu}) \simeq \frac{1}{2}\Gamma(p \to \pi^+ \bar{\nu})$, directly obtained via the isospin transformation. The neutron partial lifetime is typical at or above $10^{36}$ yr.

\section{Flavour textures of matter fields \label{sec:flavour}}

In this section we compute the proton decay branching ratios in two scenarios where Yukawa mass matrices are assumed in special textures. We show how the SUSY flipped SU(5) matches with fermion data via these scenarios. Before the discussion, we list the mass and Yukawa properties in the flipped $SU(5)$ model. 
We have three free Yukawa coupling matrices $Y_\nu$, $Y_e$ and $Y_d$ and one Majorana matrix $M_{\nu^c}$, with $Y_d$ and $M_{\nu^c}$ diagonal. The up-quark Yukawa coupling matrix is correlated with the Dirac Yukawa coupling matrix between the left- and right-handed neutrinos as $Y_u = Y_\nu^T$, and the Majorana mass matrix for light neutrinos is given by $M_\nu = - Y_\nu M_{\nu^c}^{-1} Y_\nu^T \langle h_u \rangle^2$. 
These matrices, can be diagonalised as $U_f^\dag Y_f U_f' = \hat{Y}_f \equiv {\rm diag}\{ y_{f1}, y_{f2}, y_{f3} \}$, where $f= e, u, d$ (for $f=e$, $( y_{f1}, y_{f2}, y_{f3} ) = ( y_{e}, y_{\mu}, y_{\tau} )$, etc) and $U_\nu^\dag M_\nu U_\nu^* = \hat{M}_\nu \equiv {\rm diag} \{ m_1, m_2, m_3\}$. Here, $U_l$ is the unitary matrix contributing to the proton decay ratio in the Eq.~\eqref{eq:decay_ratio}. As $Y_d$ is symmetric, $U_d' = U_d^*$ is satisfied. After the diagonalisation, we obtain the quark and lepton flavour mixing matrices as
$V_{\rm CKM} = U_u^\dag U_d$ and 
$U_{\rm PMNS} = U_l^\dag U_\nu$. 
In particular, the lepton flavour mixing matrix, i.e., the PMNS matrix, up to three unphysical phases on the left hand side, is parametrised as follows
\begin{eqnarray}
U_{\rm PMNS} = P_l
	\left(\begin{matrix}
		c^{}_{12} c^{}_{13} & s^{}_{12} c^{}_{13} &
		s^{}_{13} e^{-{ i} \delta} \cr \vspace{-0.4cm} \cr
		-s^{}_{12} c^{}_{23} - c^{}_{12}
		s^{}_{13} s^{}_{23} e^{{ i} \delta} & c^{}_{12} c^{}_{23} -
		s^{}_{12} s^{}_{13} s^{}_{23} e^{{ i} \delta} & c^{}_{13}
		s^{}_{23} \cr \vspace{-0.4cm} \cr
		s^{}_{12} s^{}_{23} - c^{}_{12} s^{}_{13} c^{}_{23}
		e^{{ i} \delta} &- c^{}_{12} s^{}_{23} - s^{}_{12} s^{}_{13}
		c^{}_{23} e^{{ i} \delta} &  c^{}_{13} c^{}_{23} \cr
	\end{matrix} \right)\, P_\nu \,,
 \label{xxx}
\end{eqnarray}
where $\theta_{ij}$ (for $ij = 12,13,23$) are three mixing angles, $\delta$ is the Dirac CP phase, $P_\nu = {\rm diag}\{1,e^{i\alpha_{21}/2},e^{i \alpha_{31}/2}\}$ is the Majorana phase matrix and $P_l = {\rm diag} \{ e^{i \beta_1}, e^{i \beta_2}, e^{i \beta_3}\}$ is a diagonal phase matrix without physical correspondence at low energy. The CKM matrix can be parameterised similarly to~(\ref{xxx}) with a $3\times 3$ matrix in the middle, involving three mixing angles $\theta_{ij}^q$ and a Dirac CP phase $\delta^q$, accompanied with two diagonal phase matrices $P_u$ and $P_d$ on both sides. However, as quarks are all Dirac fermions, the two phase matrices are unphysical at low energy. 

\begin{figure}[t!]
\centering
\includegraphics[height=.9\textwidth]{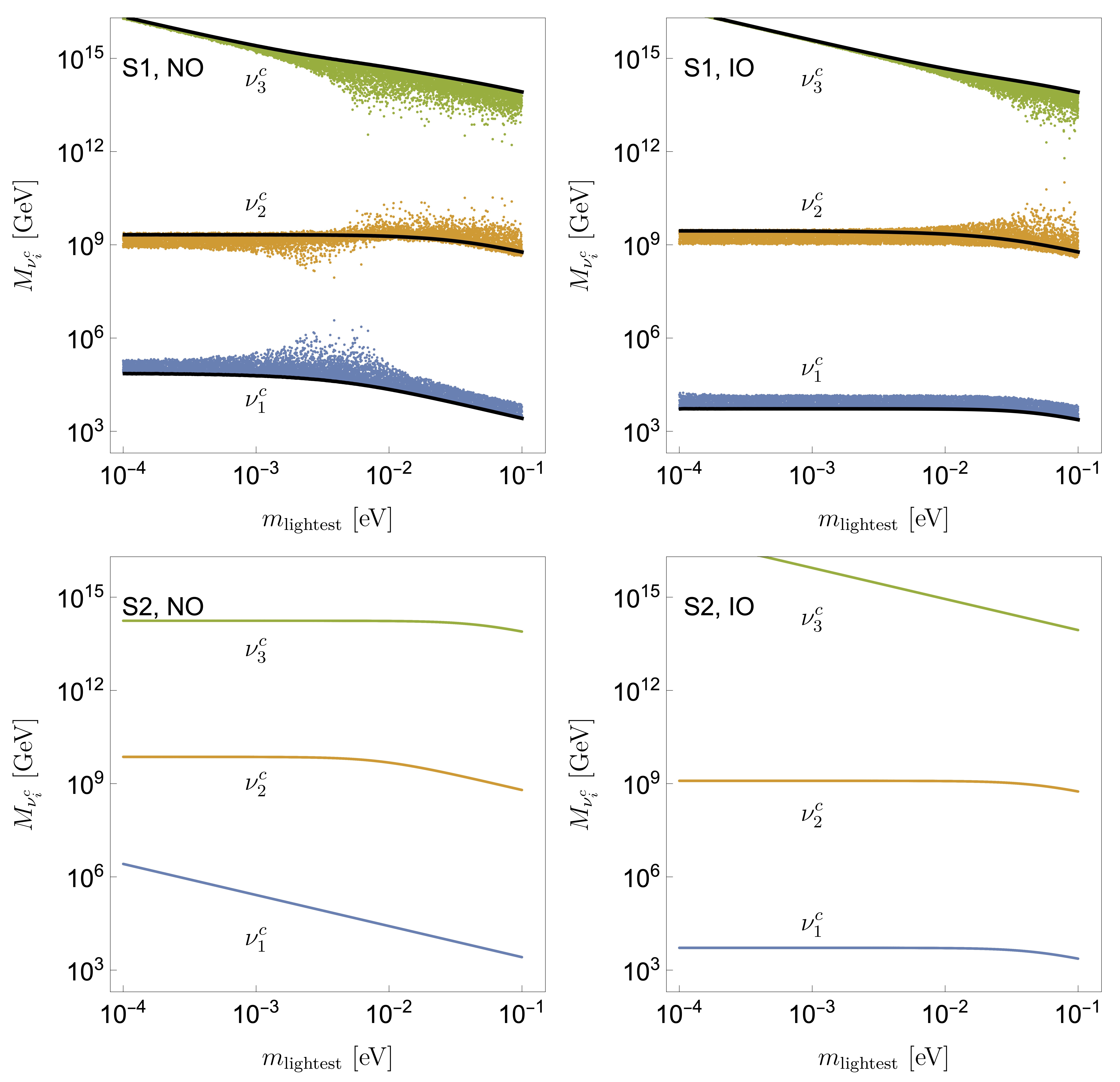}
\caption{\label{fig:RHN_mass} Prediction of RH neutrino mass spectrum vs the lightest active neutrino mass in normal ordering (NO, left panels) and inverted ordering (IO) of light neutrino masses. Black curves in the upper panels refer to all phases set to be zeros.}
\end{figure}

Without loss of generality, we make a basis rotation to the basis where $U_u = U_u'= U_l' = 1$. This is done by performing $3\times 3$ unitary transformations for $({\bf 10}, -\frac12)$, $(\bar{\bf5},\frac32)$, $({\bf1},-\frac52)$ in their flavour space, respectively. Then, we arrive at
\begin{eqnarray}
Y_u &=& Y_\nu = \hat{Y}_u\,,\nonumber\\
Y_d &=& V_{\rm CKM} \hat{Y}_d Y_{\rm CKM}^T \,, \nonumber\\
Y_l &=& U_l \, \hat{Y}_l \,, \nonumber\\
M_\nu &=& U_l U_{\rm PMNS} \hat{M}_\nu U_{\rm PMNS}^T U_l^T \,, \nonumber\\
M_{\nu^c} &=& \hat{Y}_u U_l^* U_{\rm PMNS}^* \hat{M}_\nu^{-1} U_{\rm PMNS}^\dag U_l^\dag \hat{Y}_u \langle h_u \rangle^2~.
\end{eqnarray}
In this basis, $Y_u$ and $Y_d$ are fully fixed. The rest of the Yukawa  mass matrices depend on $U_l$, the only undetermined unitary matrix involved  in their flavour structures. 
Below, we discuss two extreme cases with $U_l$ fixed.
\begin{itemize}
\item[S1)] $U_l = {\bf 1}$. This case forbids the decays $p\to K^0 \mu^+$ and $p \to K^0 \mu^+$. The branching ratios  for  the remaining channels are predicted to be
\begin{eqnarray}
{\rm BR}(p \to \pi^0 e^+) &=& 0.32\,, \nonumber\\
{\rm BR}(p \to K^0 e^+) &=& 0.006\,, \nonumber\\
{\rm BR}(p \to \pi^+ \bar{\nu}) &=& 0.68 \,.
\end{eqnarray}
The Yukawa  mass matrices are simplified to
\begin{eqnarray}
Y_l &=& \hat{Y}_l \,, \nonumber\\
M_\nu &=& U_{\rm PMNS} \hat{M}_\nu U_{\rm PMNS}^T \,, \nonumber\\
M_{\nu^c} &=& \hat{Y}_u U_{\rm PMNS}^* \hat{M}_\nu^{-1} U_{\rm PMNS}^\dag \hat{Y}_u \langle h_u \rangle^2~.
\end{eqnarray}
A typical numerical pattern of $M_{\nu^c}$ is given by
\begin{eqnarray}
|M_{\nu^c}| \simeq 
\left(
\begin{array}{ccc}
 2.60\times 10^3 & 7.09\times 10^4 & 2.32\times 10^7 \\
 7.09\times 10^4 & 5.87\times 10^8 & 1.21\times 10^{10} \\
 2.32\times 10^7 & 1.21\times 10^{10} & 8.32\times 10^{13} \\
\end{array}
\right)\,{\rm GeV}~,
\end{eqnarray}
where $m_1=0.1$~eV is assumed and phases in $P_l$ and $P_\nu$ are set at zeros. The eigenvalues for $M_{\nu^c}$ are given by
\begin{eqnarray}
(M_{\nu^c_1}, M_{\nu^c_2}, M_{\nu^c_3})  \simeq (2.61\times 10^3,5.86\times 10^8,8.32\times 10^{13})~{\rm GeV}\,.
\end{eqnarray}
We see in this case that the RH neutrinos  are very hierarchical. 
Due to the uncertainty of the complex phases, the RHN masses can still vary in a wide range, but they are in general very hierarchical. This is shown in the upper panel in Fig~\ref{fig:RHN_mass}, where both normal ordering (NO) and inverted ordering (IO) for light neutrino masses are considered. The heaviest right-handed neutrino mass varies between $10^{13}$ and $10^{16}$~GeV, depending on the mass of the lightest active neutrino mass. To satisfy the perturbativity condition, a lower bound on the $B-L$ scale  should be considered, i.e., $M_{B-L} \gtrsim M_{\nu_3^c}$. Some lower $B-L$ scale cases, e.g., $M_{B-L} = 10^{5}, 10^{9}$~GeV listed in Fig.~\ref{fig:RGE}, are not consistent with the upper bound.

\item[S2)] $U_l = U_{\rm PMNS}^\dag$.
Branching ratios of proton decay channels are numerically given by
\begin{eqnarray} \label{eq:decay_ratio_2}
{\rm BR}(p \to \pi^0 e^+) &=& 0.22\,, \nonumber\\
{\rm BR}(p \to \pi^0 \mu^+)&=& 0.080 \,, \nonumber\\
{\rm BR}(p \to K^0 e^+) &=& 0.004\,, \nonumber\\
{\rm BR}(p \to K^0 \mu^+)&=& 0.002 \,, \nonumber\\
{\rm BR}(p \to \pi^+ \bar{\nu}) &=& 0.69 \,.
\end{eqnarray}
This case leads to 
\begin{eqnarray}
Y_l &=& U_{\rm PMNS}^\dag \, \hat{Y}_l \,, \nonumber\\
M_\nu &=& \hat{M}_\nu \,, \nonumber\\
M_{\nu^c} &=& \hat{Y}_u \hat{M}_\nu^{-1} \hat{Y}_u \langle h_u \rangle^2 \,.
\end{eqnarray}
Here, the right-handed neutrino mass matrix $M_{\nu^c}$ is diagonal, with three eigenvalues have clear correlation with light neutrino masses $M_{\nu^c_1}$, $M_{\nu^c_2}$, and $M_{\nu^c_3}$ given by $m_u^2 / m_1$, $m_c^2 / m_2$, and $m_t^2 / m_3$, respectively. Taking $m_1 = 0.1$~eV in normal ordering, we obtain
\begin{eqnarray}
(M_{\nu^c_1}, M_{\nu^c_2}, M_{\nu^c_3}) 
\simeq (2.61\times 10^{3}, 6.23\times 10^{8}, 7.81 \times 10^{13})~{\rm GeV}\,.
\end{eqnarray}
A general correlation for the RHN mass spectrum with the lightest neutrino mass is given in the lower panel of Fig.~\ref{fig:RHN_mass}.
\end{itemize} 

\section{Gravitational Waves }

Recently, a series of PTA collaborations \cite{NANOGrav:2023gor,EPTA:2023fyk,Reardon:2023gzh,Xu:2023wog}
have reported  their latest data set on stochastic gravitational wave background (SGWB) emissions which provide strong evidence of  SGWB signals  with low frequences in the range of nanohertz (nHz). 		
Metastable cosmic strings can provide a potential interpretation to the observed SGWB signal \cite{NANOGrav:2023hvm}.

In the context of flipped $SU(5)$, the cosmic strings scenario is realised with the following sequence of events: First, the spontaneous symmetry breaking of $SU(5)$
to $SU(3)_c\times SU(2)_L \times U(1)_y$, achieved at a scale $M_{\rm GUT}\sim 2\times 10^{16} $ GeV, will produce heavy monopoles of immense density. They carry  $Z_3$ colour magnetic charge  and $Z_2$ electroweak magnetic charge \cite{Preskill:1984gd}, as well as minimal magnetic charge $\frac{1}{2 g_y}$ under $U(1)_y$ and $g_y$ is connected with $g_{B-L}$ via Eq.~\eqref{eq:alpha_BL}.
In order to avoid cosmological issues their density must be diminished.   
A dominant  role for their elimination is played by the inflationary scenario 
with the  inflaton  being  either a neutral singlet (such as the one appearing in the decomposition of ${\bf 24}$ (see~(\ref{W24})) or by a modulus field which  appears in the K\"ahler potential. In such a case, the scale of inflation can be around $\lesssim 10^{16}$ GeV  just below $M_{\rm GUT}$.    During the period of inflation the monopoles are diluted and their  density drops dramatically, making their current  detection impossible.

Next,  the spontaneous breaking of the $ U(1)_{\chi}$ gauge factor at the high-scale $M_{B-L}$ occurs where a network of long cosmic strings is copiously produced. According to the RGE analysis presented above,  the scale $M_{B-L}$ is somewhat lower than that of $M_{\rm GUT}$.

Then, when such cosmic strings intersect, they form string loops 
which subsequently  oscillate and  through their decay emit  Gravitational Waves. 	Superposition of GWs emitted by the various network's loops create a stochastic background in a wide frequency band. In the nHz range, it  reproduces the experimentally observed low frequency signal \cite{NANOGrav:2023hvm}.

The ratio of the GW energy density to the critical energy density $\rho_c$ in the frequency space is described by
\begin{eqnarray}
    \Omega_{\rm GW}(f) = \frac{1}{\rho_c} \frac{d\rho_{\rm GW}(f)}{d\log f}
\end{eqnarray}
It sums gravitational radiations from all strings loops in the past with redshift $z$ on the frequency considered,  
\begin{eqnarray} \label{eq:GW}
    \Omega_{\rm GW}(f)h^2 = \sum_{k=1}^{N_k} \frac{8\pi}{3} \left( \frac{h}{H_0}  (G\mu)^2 \right)^2 P_k \frac{2k}{f} \int_0^{\infty} \frac{dz}{H(z)(1+z)^6} n(l(z,k),t(z)) \,.
\end{eqnarray}
Here, $H_0 = 100 h~{\rm km/s/Mpc}$ is the Hubble expansion rate today. In the early Universe, we use the Hubble expansion rate $H(z) = H_0 \sqrt{\Omega_\Lambda + \Omega_R (1+z)^3 + \Omega_M (1+z)^4}$  in the $\Lambda$CDM model with $\Omega_R = 9.1476 \times 10^{-5}$, $\Omega_M = 0.308$ and $\Omega_\Lambda = 1 - \Omega_R - \Omega_M$, where we have ignored the small correlation due to variation of the effective d.o.f. in the thermal bath for temperature above MeV scale. $P_k = \Gamma k^{-\frac43}/\zeta (\frac43)$ (with $\Gamma \simeq 50$) describes the average gravitational  radiation power from cusps in a loop respecting to the harmonic mode $k$ in the loop oscillation \cite{Blanco-Pillado:2017oxo}. We sum the $k$ mode to a sufficiently large integer $N_k$ in our calculation. In this work, we choose $N_k = 10^5$. 
$n(l,t)$ is the loop number density distribution function for loop length between $(l, l+dl)$ at time $t$, and the loop length $l$ is correlated with the frequency today $f$ via $l(z,k) = \frac{2k}{(1+z)f}$. In the case of stable strings, the simulation result obtained by assuming Nambu-Goto (NG) strings \cite{Blanco-Pillado:2011egf,Blanco-Pillado:2013qja} gives 
\begin{eqnarray}
    n_{\rm NG}(l,t) \simeq \frac{0.18}{(l/t+\Gamma G\mu)^{\frac52} t^4} \theta(0.1 - l/t)
\end{eqnarray}
The GW spectrum predicted from stable strings depends mainly on one physical parameter $G\mu$.  It is known that such a spectrum cannot explain the NANOGrav-15 signal \cite{NANOGrav:2023hvm}. This is simply explained below: The GW signal observed in NANOGrav requires a large $G\mu$ value to enhance the GW energy density; however, such large $G\mu$ provides only a relatively flat GW spectrum in the nHz band, which cannot match with the sharper power law indicated in the NANOGrav observation.

Metastable strings have been considered as an ideal explanation to the NANOGrav hint \cite{Buchmuller:2023aus,Lazarides:2023rqf,Lazarides:2023bjd}. This scenario happens if  the string formation energy scale $\sim \sqrt{\mu}$ is not far from the monopole mass scale. In such a case, the decay channel of strings to monopole-antimonopole pairs is open. This channel does not depend on the background of monopoles. It is worth  mentioning that we do not change the timeline of the earlier history of the Universe. Namely, the $SU(5)$ is broken first, leading to the production of monopoles, which are then diluted by inflation. The process for cosmic strings forming from the $U(1)_{B-L}$ breaking is required in the end of the inflation. The low efficiency of monopole productions from string decays does not lead to cosmological problem. These monopoles do not carry unconfined fluxes after the electroweak breaking \cite{Lazarides:2023ksx}. Note that although the $SU(5)$ scale is close the mass scale of $U(1)_{B-L}$ breaking, the time scale of topological monopole production from the $SU(5)$ breaking does not have to overlap with that of cosmic string generation from $U(1)_{B-L}$ breaking. A sufficiently long time gap between the two processes allows an insertion of a period of inflation to dilute away the monopole produced in the $SU(5)$ breaking and leave cosmic strings inside the horizon \cite{Antusch:2023zjk}.

Due to the decay of the strings, the GW spectrum is suppressed in the lower frequency, leading to drop in the spectrum proportional to $f^2$. 
We follow \cite{Buchmuller:2019gfy,Buchmuller:2020lbh,Buchmuller:2021mbb,Masoud:2021prr} to formulate the GW spectrum from metastable strings.
The instability of the string is described by a decay width per unit length $\Gamma_d$ with an exponential suppression characterised by $\kappa$,
\begin{eqnarray}
    \Gamma_d = \frac{\mu}{2\pi} e^{-\pi \kappa}\,,\quad
    \sqrt{\kappa} = \frac{M_{\rm GUT}}{\alpha_5 \sqrt{\mu}} \,,
\end{eqnarray}
where the monopole mass has been approximated to $M_{\rm GUT}/\alpha_5$ by ignoring the order-one undermined factor \cite{Preskill:1984gd}. For $\sqrt{\kappa} \gtrsim 9$, a string with sub-horizon length has a lifetime $t_s \simeq 1/\sqrt{\Gamma_d}$ comparable with the lifetime of the Universe and thus is considered as a stable string. For a smaller $\sqrt{\kappa}$ value, strings and loops can decay, and an exponentially suppression factor $e^{-\Gamma_d[l t + \frac12 \Gamma G\mu (t-t_s)^2]}$ should be included in the loop number density $n(l,t)$. 

\begin{figure}[t!]
\centering
\includegraphics[height=.45\textwidth]{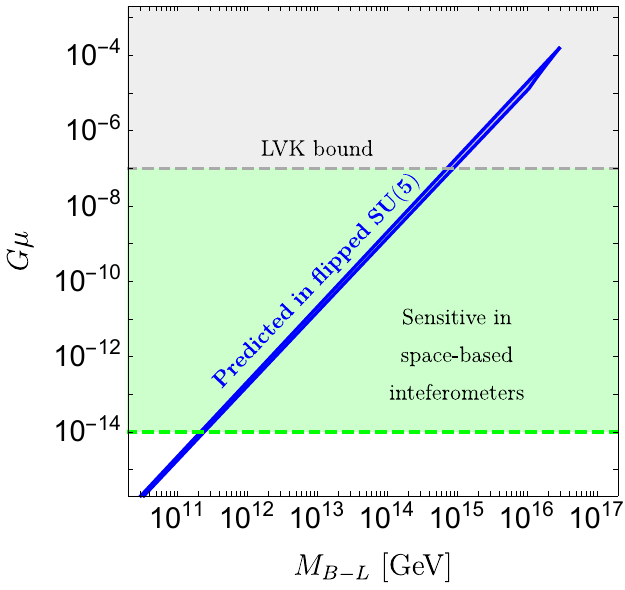}
\includegraphics[height=.45\textwidth]{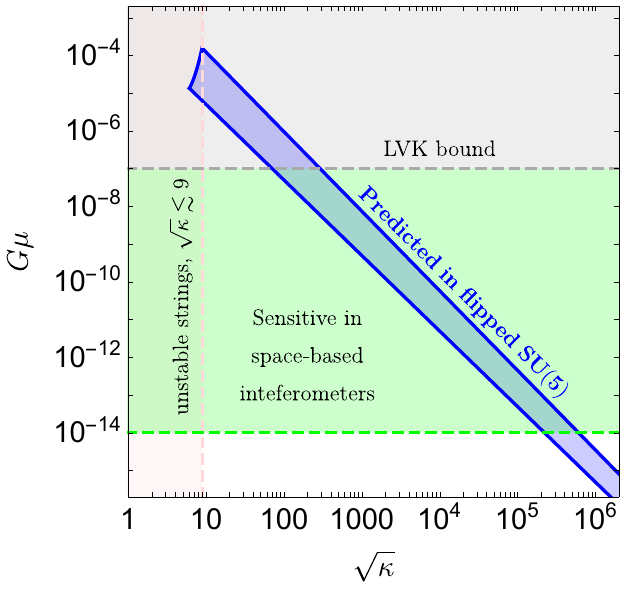}
\caption{\label{fig:Gmu_kappa} Prediction of the string tension, its correction with the scale $M_{B-L}$ and the string stability in flipped $SU(5)$. In the right panel, $G\mu \lesssim 10^{-7}$ is set by current ground-based interferometers LVK for stable strings. Space-based interferometers such as LISA, Taiji and TianQin will have the ability to test GWs for strings with $10^{-14} \lesssim G\mu \lesssim 10^{-7}$, which is indicated in the green region of the plot. $\sqrt{\kappa}$ is the parameter representing the stability of strings, and strings with $\sqrt{\kappa} \lesssim 9$ are considered as unstable strings.}
\end{figure}

In Fig.~\ref{fig:Gmu_kappa}, we show the predictions of key parameters of strings for our model. As the unification of $SU(3)_c$ and $SU(2)_L$ gauge couplings are insensitive to the $B-L$ scale. The string tension $\mu$, which is proportional to $M_{B-L}^2$, can vary in a large range up to the value referring to the GUT scale $M_{\rm GUT}$. We set a bound for $\sqrt{\kappa} \lesssim 9$ for the instability of strings. Such a small value of $\kappa$ requires energy scales of GUT breaking, inflation and $U(1)$ breaking not far away from each other, see \cite{Antusch:2023zjk} for inflationary model building in SUSY GUTs. In the right panel of Fig.~\ref{fig:Gmu_kappa}, there is a small region, with $G\mu \sim 10^{-5}$, leading to metastable strings. This is higher than the typical values discussed recently, e.g., Refs.~\cite{Lazarides:2023rqf,Buchmuller:2023aus} assumes $G\mu$ located somewhere around $10^{-6}$ or $10^{-7}$. 

In our case, in order to generate GWs consistent with NANOGrav 15 from oscillation of these heavier strings, inflationary period to dilute some of GWs at higher frequency is necessary, such that the LIGO/Virgo/KAGRA (LVK) bound $G\mu \lesssim 10^{-7}$ can be avoided. 
Including the influence of inflation is natural in the flipped $SU(5)$ as the energy scale of string generation scale is very close to the monopole scale. Thus, the string network is likely to form just in the last stage before the inflation is fully complete. Due to the influence of inflation, the string loops generated in the string network are firstly inflated and then re-enter the horizon at a certain time $t(z_{\rm inf})$ respecting to the redshift $z_{\rm inf}$. A simple treatment to including the influence of inflation is introducing a step function $\theta(t - t(z_{\rm inf}))$. 

\begin{figure}[t!]
\centering
\includegraphics[width=.85\textwidth]{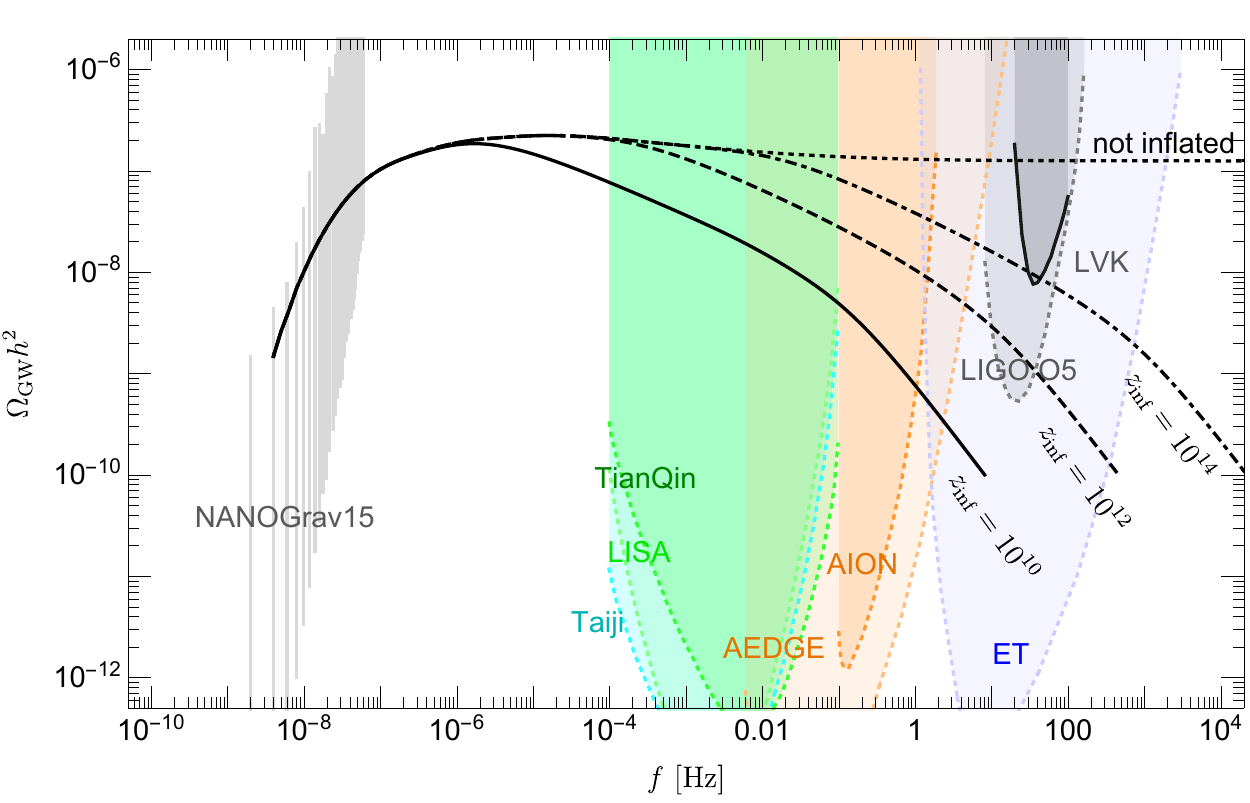}
\caption{\label{fig:GW} Benchmark of GW spectrums predicted in SUSY flipped $SU(5)$. $G\mu = 10^{-5}$ is used. $\sqrt{\kappa} = 7.8$ is tuned to fit the signal of NANOGrav 15 data in the nHz region. Three curves of GW spectrum with strings generated during the end of inflation are shown with strings re-entering the horizon at the redshift $z_{\rm inf} = 10^{10}$, $10^{12}$ and $10^{14}$, respectively. The $k$ mode is summed up to $N_k = 10^5$. GW generated fully after inflation is shown in comparison. }
\end{figure}

Including  all these effects together, we obtain the modified loop number density distribution function as
\begin{eqnarray}
    n(l,t) = n_{\rm NG}(l,t) \, e^{-\Gamma_d[l t + \frac12 \Gamma G\mu (t-t_s)^2]} \, \theta(t - t(z_{\rm inf}))
\end{eqnarray}
Taking it into Eq.~\eqref{fig:GW}, we are able to obtain GW spectrums compatible with NANOGrav signal and consistent with the LVK bound. 
We show three benchmarks of GW spectra  in Fig.~\ref{fig:GW}, with $z_{\rm inf} = 10^{10}$, $10^{12}$ and $10^{14}$ assumed, referring to the time for strings re-enter the horizon at $t_{\rm inf} = 0.26$, $3.6\times 10^{-5}$ and $3.8 \times 10^{-9}$ second, respectively. Another curve of GW spectrum with strings not inflated by the inflation is shown for comparison, which conflicts with the LVK bound. 

%%%%%%%%%%%%%%%%%%%%%%%%%%%%%%%%%%%%%%%%%%%%%%%%%%%%%%%%%%%%%%%%%%%%%%%%
%%%%%%%%%%%%%%%%%%%%%%%%%%%%%%%%%%%%%%%%%%%%%%%%%%%%%%%%%%%%%%%%%%%%%%%%

\section{Conclusion}

In this paper we have focused on the 
experimental tests of SUSY flipped $SU(5)$, including unification, proton decay, fermion masses and gravitational waves which arise from cosmic string loops associated with the breaking of the high energy $U(1)_{B-L}$.

We have presented the spectrum of the model and described the  breaking  pattern of the flipped $SU(5)\times U(1)_{\chi}$ symmetry 	
via a sequence of scales.
We performed a two-loop renormalisation group analysis for the gauge coupling unification of the $SU(5)$ factor.
We found that the SUSY flipped $SU(5)$ model predicts a high GUT scale $M_{\rm GUT} > 10^{16}$~GeV. 
We also investigated  the restrictions on the $M_{B-L}$ scale which is associated with the $U(1)_{\chi}$ breaking scale.
We found that the $M_{B-L}$ scale can vary in a broad region with negligible or little effect on the value of $M_{\rm GUT}$. 

Proton decay in this model is induced by dimension-6 operators only.
The dimension-5 operator induced by SUSY contribution is suppressed due to the missing partner mechanism. 
We found that the partial decay width $p \to \pi^0 e^+$ is highly suppressed, being at least one order of magnitude lower than the future Hyper-K sensitivity. This suppression originates for two reasons: one due to the high GUT scale, the other because there are fewer dimension-6 operators compared with the traditional $SU(5)$ or $SO(10)$ models. 
Nevertheless we calculated all the decay channels of protons. 
The dominant channel is not $p \to \pi^0 e^+$ but $p \to \pi^+ \bar{\nu}$. The GUT scale is correlated with the SUSY breaking scale, but is insensitive to the $B-L$ breaking scale. 

We also studied fermion (including neutrino) masses and mixings which can also influence proton decay.
We presented two scenario of flavour textures to check the consistency of the results with fermion masses and mixing. 

The $B-L$ gauge breaking leads to the generation of cosmic strings. The $B-L$ scale here is not constrained by gauge coupling unification. 
If this  scale is very close to that of GUT breaking, strings can be unstable due to the decay to monopole-antimonopole pair. Such metastable strings can be used to explain the NANOGrav signals
of stochastic gravitational wave background, which may be interpreted here as resulting from the decay of metastable cosmic strings.

The present study complements previous recent studies where such testable predictions have been studied in the framework of $SO(10)$. Unlike the case of SUSY $SO(10)$ the SUSY breaking scale and the $U(1)_{B-L}$ breaking scales are essentially free parameters in flipped $SU(5)$, allowing for a wide range of gravitational wave signatures. This freedom allows the NANOGrav data to be fitted more easily in SUSY flipped $SU(5)$ than in SUSY $SO(10)$, for high $B-L$ breaking scales where unstable strings are possible, due to their decay into monopole-antimonopole pairs, without violating high frequency LIGO bounds. By contrast, in SUSY $SO(10)$, proton decay places a severe constraint on this scenario.

\section*{Acknowledgement}

This work is supported by the STFC Consolidated Grant ST/L000296/1 and the European Union’s Horizon 2020 Research and Innovation programme under Marie Sklodowska-Curie grant agreement HIDDeN European ITN project (H2020-MSCA-ITN-2019//860881-HIDDeN) (S.F.K.), the Hellenic Foundation for Research and Innovation (H.F.R.I.) under the “First Call for H.F.R.I. Research Projects to support Faculty Members and Researchers and the procurement of high-cost research equipment grant” (Project Number: 2251) (G.K.L.), and National Natural Science Foundation of China (NSFC) under Grants No. 12205064 and Zhejiang Provincial Natural Science Foundation of China under Grant No. LDQ24A050002 (Y.L.Z.).
G.K.L. would like to thank the staff of Nuclear and Particle Physics, of the Faculty of Physics of NKUA in Athens, for the kind hospitality where part of this work has been carried out. Y.L.Z would like to thank Y.L. Wu for useful discussions.

%%%%%%%%%%%%%%%%%%%%%%%%%%%%%%%%%%%%%%%%%%%%%%%%%%%%%%%%%%%%%%%%%%%%%%%%
%%%%%%%%%%%%%%%%%%%%%%%%%%%%%%%%%%%%%%%%%%%%%%%%%%%%%%%%%%%%%%%%%%%%%%%%
%%%%%%%%%%%%%%%%%%%%%%%%%%%%%%%%%%%%%%%%%%%%%%%%%%%%%%%%%%%%%%%%%%%%%%%%

\end{document}